# Electric Field Control of Three-Dimensional Vortex States in Core-Shell Ferroelectric Nanoparticles

Anna N. Morozovska [1*], Eugene A. Eliseev[2], Riccardo Hertel[3†], Yevhen M. Fomichov[4], Viktoriia Tulaidan[5], Victor Yu. Reshetnyak[5‡], and Dean R. Evans[6§]

[1] Institute of Physics, National Academy of Sciences of Ukraine,

46, pr. Nauky, 03028 Kyiv, Ukraine

[2] Institute for Problems of Materials Science, National Academy of Sciences of Ukraine,

Krjijanovskogo 3, 03142 Kyiv, Ukraine

[3] Université de Strasbourg, CNRS, Institut de Physique et Chimie des Matériaux de Strasbourg,

UMR 7504, 67000 Strasbourg, France

[4] Charles University in Prague, Faculty of Mathematics and Physics,

V Holešovičkach 2, Prague 8, 180 00, Czech Republic

[5] Taras Shevchenko National University of Kyiv, Volodymyrska Street 64, Kyiv, 01601, Ukraine

[6] Air Force Research Laboratory, Materials and Manufacturing Directorate, Wright-Patterson Air Force Base, Ohio, 45433, USA

\*     Corresponding author 1: anna.n.morozovska@gmail.com

†     Corresponding author 2: riccardo.hertel@ipcms.unistra.fr

‡     Corresponding author 3: victor.reshetnyak@gmail.com

§     Corresponding author 4: dean.evans@us.af.mil

## Abstract

The fundamental question whether the structure of curled topological states, such as ferroelectric vortices, can be controlled by the application of an irrotational electric field is open. In this work, we studied the influence of irrotational external electric fields on the formation, evolution, and relaxation of ferroelectric vortices in spherical nanoparticles. In the framework of the Landau-Ginzburg-Devonshire approach coupled with electrostatic equations, we performed finite element modeling of the polarization behavior in a ferroelectric barium titanate core covered with a "tunable" paraelectric strontium titanate shell placed in a polymer or liquid medium.

A stable two-dimensional vortex is formed in the core after a zero-field relaxation of an initial random or poly-domain distribution of the polarization, where the vortex axis is directed along one of the core crystallographic axes. Subsequently, sinusoidal pulses of a homogeneous electric field with variable period, strength, and direction are applied. The field-induced changes of the vortex structure consist in the appearance of an axial kernel in the form of a prolate nanodomain, the kernel growth, an increasing orientation of the polarization along the field, and the onset of a single-domain state. We introduced the term "kernel" to name the prolate nanodomain developed near the vortex axis and polarized perpendicular to the vortex plane. In ferromagnetism, this region is generally known as the vortex core.

Unexpectedly, the in-field evolution of the polarization includes the formation of Bloch point structures, located at two diametrically opposite positions near the core surface. After removal of the electric field, the vortex recovers spontaneously; but its structure, axis orientation, and vorticity can be different from the initial state. As a rule, the final state is a stable three-dimensional polarization vortex with an axial dipolar kernel, which has a lower energy compared to the initial purely azimuthal vortex. The nature of this counterintuitive result is that the gradient energy of the axial vortex without a kernel is significantly higher, while the formation of a vortex kernel only leads to a smaller increase of the depolarization energy.

The analysis of the torque and electrostatic forces acting on the core-shell nanoparticle in an irrotational electric field showed that the torque acting on the vortex with a kernel tends to rotate the nanoparticle in such way that the vortex axis becomes parallel to the field direction. The vortex (with or without a kernel) is electrostatically neutral, and therefore the force acting on the nanoparticle is absent for a homogeneous electric field, and nonzero for the field with a strong spatial gradient.

The vortex states with a kernel possess a manifold degeneracy, appearing from three equiprobable directions of vortex axis, clockwise and counterclockwise directions of polarization rotation along the vortex axis, and two polarization directions in the kernel. This multitude of the vortex states in a single core are promising for applications of core-shell nanoparticles and their ensembles as multi-bit memory and related logic units. The rotation of a vortex kernel over a sphere, possible for the core-shell nanoparticles in a soft matter medium with controllable viscosity, may be used to imitate qubit features.

# I. INTRODUCTION

Nanosized ferroelectrics attract permanent attention of researchers as unique model systems for fundamental studies of polar surface properties, various screening mechanisms of spontaneous polarization by free carriers, and often the emergence of versatile multi-domain states with complex topology of electric dipoles [1, 2, 3, 4]. Many experimental and theoretical researchers have tried to answer the question whether or not complex topological states, such as flux-closure domains or polarization vortices and skyrmions, can exist in bulk and nanosized ferroelectrics, and whether one can control these topological states by an external stimulus (see e.g. Refs. [5, 6, 7, 8, 9] and citations therein).

It follows from many theoretical studies that a homogeneous electric field alone is practically incapable of controlling the vortex polarization of nanosized ferroelectrics, and that additional features are required, such as particle curvature or specific defects [10, 11]. In contrast, it has been shown theoretically that "curled" electric fields [12], or asymmetric mechanical fields [9, 13], which are very hard to realize experimentally, can induce a switching of the vortex polarization.

However, there are several experimental studies on the behavior of vortex polarization in nanosized ferroelectrics under the application of an irrotational electric field [5, 14]. Specifically, Rodriguez et al. [15] studied two-dimensional arrays of ferroelectric lead zirconate titanate nanodots with the help of a strongly inhomogeneous electric field created by the probe of a piezoresponse force microscope (**PFM**), and observed the presence of a quasi-toroidal polarization ordering. Karpov et al. [16] used Bragg coherent diffractive imaging of a single $BaTiO_3$ nanoparticle in a composite "polymer/ferroelectric" capacitor to study the behavior of a three-dimensional vortex. They revealed a mobile vortex kernel exhibiting a reversible hysteretic transformation path under the influence of an external electric field.

A much greater number of studies is devoted to the phase-field modeling of polarization vortices in nanosized ferroelectrics and their reaction to an external stimulus using a continuum phenomenological Landau-Ginzburg-Devonshire (**LGD**) approach combined with electrostatic equations (see Ref. [17] and refs therein). For instance, Chen and Zheng [13] performed phase-field modeling of the evolution of vortex domain structures in ferroelectric nanodots under asymmetric elastic strains induced by the substrate, dislocations, and local clamping forces. They realized single-vortex switching using a homogeneous electric field by controlling the flow direction of the dominant dipole region induced by asymmetric mechanical strains. Wu et al. [18] and Xiong et al. [19] demonstrated that surface charge screening in combination with temperature changes can provide an efficient way to gain control of a vortex domain structure in ferroelectric nanodots. Mangeri et al. [20] simulated the behavior of the polarization in isolated spherical $PbTiO_3$ and $BaTiO_3$ nanoparticles embedded in a dielectric matrix, and showed that the vortex-like polarization topology is strongly affected by the particle diameter, as well as by the choice of inclusion and matrix materials. Mangeri et al. [21] then proposed different ways for the electromechanical control of polarization vortex ordering in interacting ferroelectric-dielectric dimers. Pitike et al. [22] modeled ferroelectric nanoparticles and found that the critical particle sizes of the texture instabilities were strongly dependent on the particle shape, with octahedral particles undergoing transitions at much larger volumes, compared to cubic particles. Zhu et al. [23] modeled polar properties of ferroelectric nanoparticles with different shapes and sizes, and it was shown that the process of polarization switching occurs via an emergence of intermediate phases that involve an appreciable amount of vorticity.

It should be underlined that the electric field control of a nanosized ferroelectric polarization is of particular interest due to its perspective applications in modern electrocaloric (**EC**) convertors working around phase transition temperatures [24, 25, 26]. Using an LGD approach and numerical modeling, Li et al. [27], Zeng et al. [28], and Wang et al. [29] calculated the EC properties of ferroelectric nanoparticles with complicated vortex-like domain structures. Chen and Fang [30] studied the EC effect in barium titanate nanoparticles with vortex polarization using a core–shell model.

To the best of our knowledge, existing theoretical papers (cited above and many others) did not consider analytically the possibility of *controlling* the vortex polarization in nanosized ferroelectrics by irrotational electric fields, which can be easily created.

Motivated to fill this gap in knowledge, we simulate numerically and describe analytically the formation of polarization vortices in spherical nanoparticles consisting of a ferroelectric core covered with a paraelectric shell, and analyze the vortex behavior in irrotational electric fields. We calculate analytically the torque and the electrostatic force acting on the core-shell nanoparticle placed in a liquid (or viscous) cell in the presence of a homogeneous electric field or its spatial gradient. The controllability feature of the nanoparticle's core polarization by the shell screening in combination with external irrotational electric fields looks very attractive for new applications, such as multi-bit memory, qubit simulation, and logic units.

The remainder of the paper has the following structure. The formulation of the problem is presented in **Section II**, which contains the method description and calculation details. Results of finite element modeling and their analytical description are presented in **Section III**, where special attention is paid to the evolution of the polarization under an applied voltage, the structure of the initial and final polarization states, and the polarization behavior in an external electric field. The torque and electrostatic forces acting on the core-shell nanoparticle placed in homogeneous and inhomogeneous (gradient) electric fields are analyzed in **Section IV**. Possible applications of core-shell nanoparticles and their ensembles as multi-bit memory and related multi-value logic units, as well as speculations about their ability to imitate qubit features at room temperature are discussed in **Section V.** Obtained results are summarized in **Section VI.**

## II. FORMULATION OF THE PROBLEM

**A. Method.** We use the LGD approach combined with electrostatic equations, because this allows us to establish the physical origin of anomalies in the phase diagrams, to determine polar and dielectric properties of ferroelectric nanoparticles [31, 32], and to calculate the changes of their domain structure morphology with size reduction [33, 34]. The LGD methodology further allows the incorporation of various physical mechanisms ruling the size effects, such as surface bond contraction [35, 36], intrinsic surface stresses and strains [37, 38, 39], correlation effects, and depolarization fields originating from the incomplete screening of the polarization [40].

We performed finite element modeling (**FEM**) of the polarization and electric field evolution in a 10 nm barium titanate (BaTiO$_3$) nanoparticle covered with a 4 nm thick "tunable" strontium titanate (SrTiO$_3$) shell with a ultra-high temperature-dependent relative dielectric permittivity $\varepsilon_S \sim (3000 - 300)$, which is placed in a polymer or liquid medium with a relative dielectric permittivity $\varepsilon_e \sim 10$. The main role of the paraelectric shell is to provide an effective tunable screening of the core polarization [41]. The core size is typical of experimental values [42, 43, 44, 45].

The mathematical formulation of the problem, consisting in the electrostatic equations and time-dependent LGD equations with boundary conditions, is given in detail in Ref. [41], and it is repeated in **Appendix A** allowing for the application of a voltage to the capacitor electrodes located at the boundaries of the computational region. Parameters of the core and shell materials used in FEM are provided in **Table AI.** The sizes and screening conditions are chosen in such a way that the nanoparticle core is in a stable orthorhombic ferroelectric phase with a vortex polarization in the absence of an external electric field. The vortex polarization exists over a wide temperature range $(250 - 350)$ K [41].

**B. Calculation details.** The simulated model system is shown in **Fig. 1(a).** The upper electrode can be biased and the bottom electrode is grounded. These two electrodes can rotate around the core-shell nanoparticle, remaining parallel to each other, as shown in **Fig. 1(b)**. During the rotation, the crystallographic axes $\{x, y, z\}$ of the particle core obviously remain unchanged, but the surrounding medium rotates around the particle. This rotation is characterized by angles $\theta$ and $\psi$ (taken in a spherical coordinate system). The size of the upper electrode is varied to model the inhomogeneous electric field [see **Fig. 1(c)**]. Note that a strong field gradient was used in experiments on nanoparticle harvesting [46, 47] and PFM studies [15].

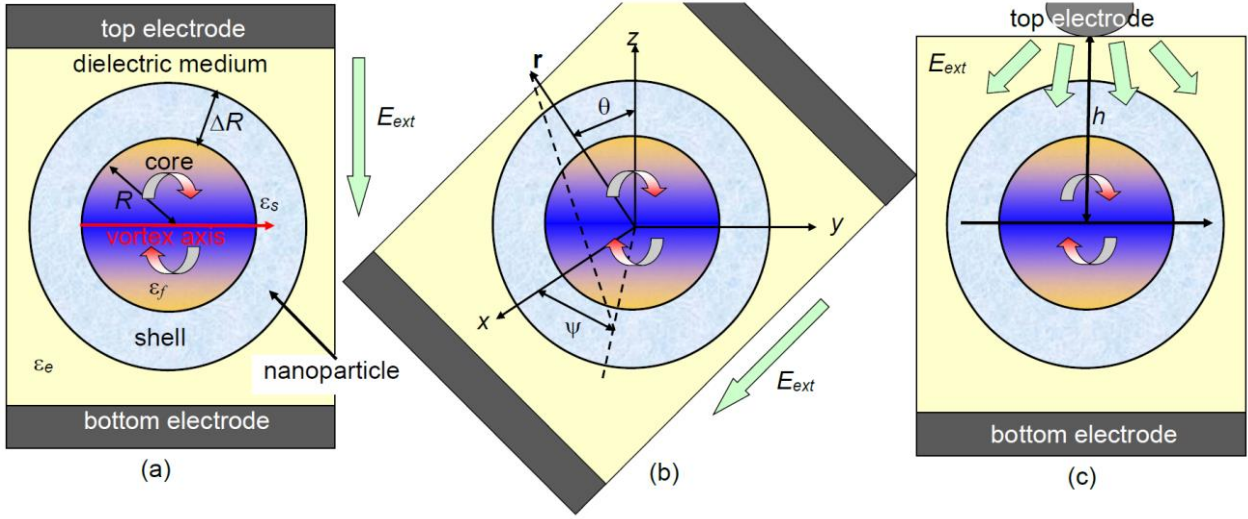

**FIGURE 1. (a)** A spherical ferroelectric nanoparticle (core) of radius *R,* covered with a paraelectric layer (shell) of thickness Δ*R,* placed in an isotropic dielectric effective medium. **(b)** Rotating electrodes around the core-shell nanoparticle. The crystallographic axes {*x, y, z*} of the particle core remain unchanged during the rotation characterized by angles θ and ψ, taken in a spherical coordinate system. **(c)** The size of the top electrode is varied, allowing for the modelling of inhomogeneous electric fields in the computational region.

Hence, FEM was carried out according to the following scheme:

1. The position of the electrodes is defined by the angles θ and ψ.

2. The voltage between the electrodes is absent in the time interval $-500\ \tau_K$ to $0\ \tau_K$ (see **Figs. 2,** left part), where the time is measured in units of the Landau-Khalatnikov relaxation time $\tau_K$ [48]. The time $\tau_K$ can vary in a relatively wide interval, from $(10^{-9} - 10^{-6})$ seconds, far from the ferroelectric phase transition, to much higher values approaching the transition that corresponds to the critical slowing down of the ferroelectric response. The time $500\ \tau_K$ appeared to be long enough for the formation of a stable single vortex from the initial random distribution of polarization. So, the characteristic time scale of "off-field relaxation" is of the order of $\sim 100\ \tau_K$.

3. A quasi-static sinusoidal voltage pulse with an amplitude of 20 V is applied during different periods. **Figure 2** shows a comparison between the two cases of sinusoidal pulses with different periods $T_p$. The results show that the particle polarization is in a dynamic regime at $T_p = 10^3\ \tau_K$ [**Fig. 2(a),** middle part], where changes in the polarization structure lag behind the applied field, and changes in the polarization approach the quasi-static regime at $T_p = 10^4\ \tau_K T$ [**Fig. 2(b),** middle part]. Note that the times $(10 - 100)\tau_K$ appear to be long enough for the system's relaxation in the electric field, and that the "in-field" relaxation time reduces with increasing voltage. Voltages below a certain critical value do not destroy the vortices, indicating the stability of these structures. Note that the relaxation time is weakly dependent on the orientation of the external field with respect to the crystallographic axes.

4. The field is absent again for $t > T_p$, and the system slowly relaxes to a new equilibrium state in the time interval $T_p < t < 2T_p$, which corresponds to a rotated and distorted vortex (see **Figs. 2**, right part). Note that the contrast in the images on the upper right indicates the appearance of a nonzero dipole moment at zero field. This dipole moment, which originates from the vortex kernel, can provide a means to manipulate the vortex orientation by an external field, as discussed in more detail later.

5. The size of the top electrode is varied to model the inhomogeneous electric field.

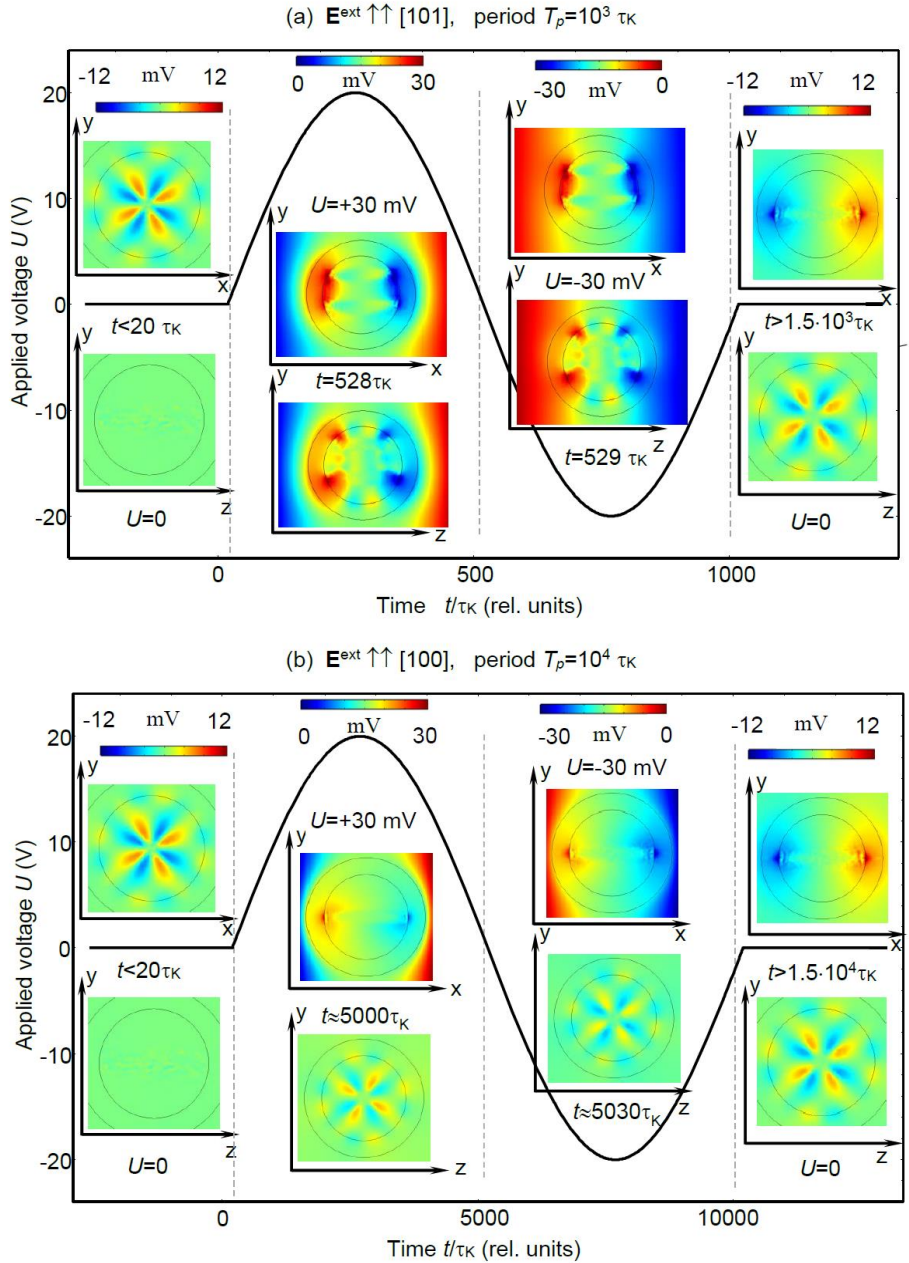

**FIGURE 2.** Evolution of the applied voltage $U$ and electric potential φ in the region of the core-shell nanoparticle. Color images show characteristic distributions of the electric potential φ inside and around the nanoparticle at different times in the units of Landau-Khalatnikov relaxation time $\tau_K$. **(a)** The vortex axis

coincides with the crystallographic direction [001] before the voltage pulse, while its axis is oriented along the [100] direction long after the pulse ending. The amplitude of the applied sinusoidal voltage pulse is 20 V, the pulse duration is $10^3 \tau_K$ **(a)** and $10^4 \tau_K$ **(b)**. The direction of the external field is [101] **(a)** and [100] **(b)**, which corresponds to the angles $\psi = 0^o$, $\theta = 45^o$ **(a)** and $\theta = 0^o$ **(b)**. Particle radius $R = 10$ nm, shell thickness $\Delta R = 4$ nm, and temperature $T$=298 K; BaTiO$_3$ core and SrTiO$_3$ shell parameters are listed in **Table AI.**

## III. FEM RESULTS AND THEIR ANALYTICAL DESCRIPTION

It is shown [41] that the vortex polarization in spherical nanoparticles with radius 10 nm or greater is stable over a wide temperature range $(250 - 350)$ K. Our FEM simulations have shown that a stable vortex is formed from various initial distributions of polarization (using either random "seeding", or different poly-domain initial conditions). The vortex axis is directed along one of the crystallographic axes, since the ferroelectric anisotropy energy of the core is minimal for these directions. After the vortex formation, an external electric field is applied to the nanoparticle, and its strength and direction with respect to the vortex axis are varied. We investigate whether the vortex polarization of a nanoparticle can be controlled by irrotational (homogeneous or gradient-type) electric fields. BaTiO$_3$ core and SrTiO$_3$ shell LGD parameters, collected from Refs. [49, 50, 51, 52, 53, 54], are listed in **Table AI**, **Appendix A.**

Our simulations demonstrate that the minimal external field required to change the vortex structure is about $(15 - 30)$ mV/nm in a quasi-static case $(T_p > 10^4 \tau_K)$. The fields must exceed the maximal depolarization electric field, that is about 15 mV/nm in the equilibrium state. Significantly higher fields $\sim (100 - 150)$ mV/nm, which are still smaller than the thermodynamic coercive field $\sim$ 250 mV/nm, are required to rotate the polarization vortex axis, or to destroy the vortex completely. The discrepancy between the calculated thermodynamic coercive field ($\sim$ 250 mV/nm) and the experimentally observed coercive field of bulk BaTiO$_3$ ($\sim$ 0.1 mV/nm) may originate from the influence of intrinsic charged defects, which we do not consider. All of these fields are significantly smaller than the breakdown field of a surrounding liquid medium or polymer $\sim (500 - 700)$ mV/nm [55].

### A. Evolution of the Polarization Under an Applied Voltage

**Figure 3** shows a typical example of the polarization components' evolution under a sinusoidal pulse of an applied voltage with a duration of $10^3 \tau_K$. The yz and xy cross-sections correspond to various snapshots taken at different delay times after the application of the external field, measured in $\tau_K$ units. A stable single vortex state with energy $G_1 \approx - 8.967 \times 10^{-18}$ J is formed from any type of polydomain (or random) polarization distribution at $t = 0$, but the formation time significantly depends on its specific form. It is greatest for a small random distribution (more than 200 $\tau_K$), and becomes considerably smaller for a closure-type polydomain polarization. It appears that the fastest formation of a vortex (about 20 $\tau_K$) occurs when the initial configuration of the polarization is a superposition of the four Kittel-type flux-closure domains and a small random polarization (it is shown in **Fig. 3** for $t$=0).

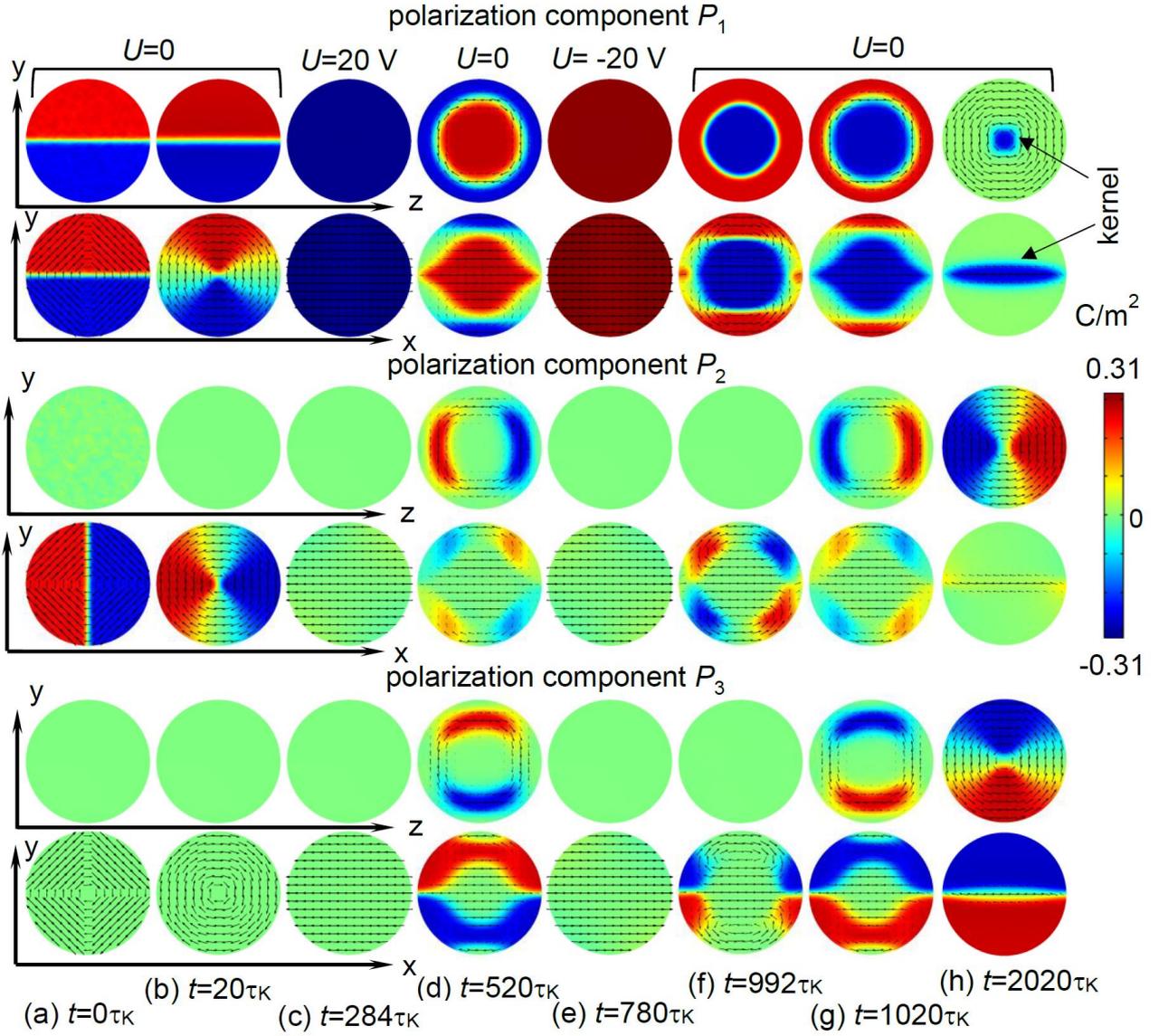

**FIGURE 3.** Temporal evolution of the polarization components' distribution in the yz and xy cross-sections of the BaTiO$_3$ nanoparticle covered with a SrTiO$_3$ shell. Several moments of dimensionless time $t$ (in the units of $\tau_K$) are shown in plots **(a) − (h)**. Black arrows indicate polarization direction. The direction of the external field is [100]. The vortex axis coincides with the crystallographic direction [001] before the voltage pulse, and is aligned with the [100] direction long after the voltage removal. The sinusoidal pulse duration is $10^3 \tau_K$. Other parameters are the same as in **Fig. 2(a)**.

We chose a situation where the vortex axis coincides with the [001] crystallographic direction before applying the voltage pulse, at $t$<0. The sinusoidal voltage pulse is shown in **Fig. 2(a)**. After the voltage is applied at $t>0$, the electrode position is defined by the angles $\theta = 90^o$ and $\psi = 0^o$, which corresponds to an external field directed along the [100] crystallographic axis. The vortex in-field behavior can be described as follows: while the voltage amplitude remained below a critical value, only slight and gradual changes occur, but as the voltage increases it eventually destroys the vortex resulting in a homogeneously polarized state for sufficiently high voltages. Then, the homogeneous state splits into a double vortex state as the voltage sign changes. After the voltage is removed, the polarization relaxation leads to the formation of a new single vortex with a lower energy $G_0 \approx - 8.978 \times 10^{-18}$ J and a rotated axis that coincides with the field direction [100]. The fully relaxed vortex structure contains a "**kernel**", that resembles a nanodomain in the form of the prolate ellipsoid of width 6.4 nm and length 20 nm, with an almost homogeneous polarization along the vortex axis [see two top images in **Fig. 3(h)**]. This ellipsoidal nanodomain in the center of the ferroelectric vortex has a shape that is similar to the characteristic vortex kernel (or core), which appears in the case of *ferromagnetic* three-dimensional vortex structures (see, e.g., chapter 3.6. of textbook [56] and Ref. [57]). The prolate shape of the vortex kernel, known from ferromagnetic vortices, is due to electrostatic reasons. By decreasing the radius of the kernel near the surface, the system reduces the amount of electrostatic surface charges $P_n$, and thereby lowers the dipole moment and depolarization energy. Note, that we use the term "kernel" to describe the region of the vortex near its axis that is polarized perpendicular to the vortex plane. In ferromagnetism, this region is generally known as the vortex "core" [58, 59]. We use "kernel" in this paper in order to avoid ambiguities in the general context of core-shell particles.

When the electric field is directed, e.g., along the [111] axis, the "relaxed" vortex axis can lie along any of three possible directions ([001], [010] or the initial [100]), depending on the computational fluctuations. Thus, there are three equilibrium directions of the vortex axis, between which the vortex can be switched via the application and subsequent removal of an external electric field. Each direction and each vortex circulation direction (clockwise or anti-clockwise) has three local minima, namely a simple vortex (**excited state "0"**) and two vortex states with a kernel having mutually opposite polarization (**ground states "±1"**). Actually, for any given sense of rotation (vorticity), the state "±1" is two-fold degenerate, as there are two equivalent states that differ by the sign of the polarization within the kernel. These types of the lowest states "0" and "±1" have a relatively small energy difference $\Delta G \approx + 1.2 \times 10^{-20}$ J, corresponding to the barrier height 2.9 $k_B T$ at 298 K and a relative energy difference of only about 0.1%. The formation of the state "±1" with a kernel can be attributed to a significant reduction of the gradient energy contribution, since an empty vortex axis has a higher density of gradient energy near the central axis in the state "0". The density becomes significantly smaller in the states "±1", when the "empty" region surrounding the vortex axis transforms into a prolate nanodomain. On the other hand, the kernel formation leads to an increase of the depolarization field energy. The difference in the free energies of the states "0" and "±1" suggests that, for this specific sample, the decrease of gradient energy of the almost homogeneously polarized kernel is partially compensated by the increase of depolarization energy that arises from the kernel dipole moment.

Given the relative proximity of the simple vortex energy $G_1$ and the energy $G_0$ of a vortex with a kernel, the polarization of a nanoparticle core can pass from one local minimum to another under certain external conditions. By placing a particle with a simple vortex polarization in an external electric field, which has a nonzero average over the particle volume, it turns a simple vortex into a vortex with a kernel regardless of the field direction, as can be seen from **Fig. 3**. The magnitude of the field may be small, but it is sufficient to overcome the thermodynamic barrier $\Delta G / k_B T$. The nanoparticle core becomes completely or partially single-domain in larger fields with a nonzero average.

In view of this result, we can expect that a simple vortex state "0" is a metastable state, i.e. that it represents an energetic local minimum. This raises the question whether the state "0" might be an unstable minimum, i.e. a saddle point, which is only formed in perfectly random (or polydomain) conditions of the simulation. A saddle point could become unstable by applying an arbitrarily small field, breaking the symmetry along the axis. However, we repeatedly found in our simulations that a simple vortex is formed spontaneously (at $\mathbf{E}^{ext}$=0) from any small random distribution of polarization, which has a certain (but rather small) degree of asymmetry [41]. These trial simulations thus indicate that state "0" is a real local minimum. We also found that in the simulations it is possible to transform a vortex with an axial kernel into a simple vortex. This, however, requires the application of a random electric field, which has a zero average over the particle volume and an amplitude that is sufficient to overcome the barrier.

The polarization of a two-fold degenerate kernel state and its quasi-static evolution at small voltages are shown in **Fig. 4.** Polarization profiles are calculated for a long period voltage pulse, $T_p = 10^4 \tau_K$. Two perpendicular line scans, $x = y = 0$ and $y = z = 0$, are shown for several small voltages: $U \geq 0$ in **Figs. 4(a, b)** correspond to the upper part of the pulse in **Fig. 2(b)**, and $U \leq 0$ in **Figs. 4(c, b)** correspond to the lower part of the pulse in **Fig. 2(b)**. Symbols are the FEM results. Solid black curves were calculated for $U = 0$ using the analytical dependence $P_1(x, y = 0, z) = P_0 + P_d \tanh\left(\sqrt{\frac{x^2}{R_x^2} + \frac{z^2}{R_z^2} + d^2} - L_d\right)$ with parameters $R_x = 1.5$ nm, $R_z = 0.8$ nm, $d = 1.5$, $L_d = 2.5$, $P_0 = -0.105$ C/m$^2$, and $P_d = 0.160$ C/m$^2$ [for **Fig. 4(a,b)**], or $P_0 = +0.105$ C/m$^2$ and $P_d = -0.160$ C/m$^2$ [for **Fig. 4(c,d)**].

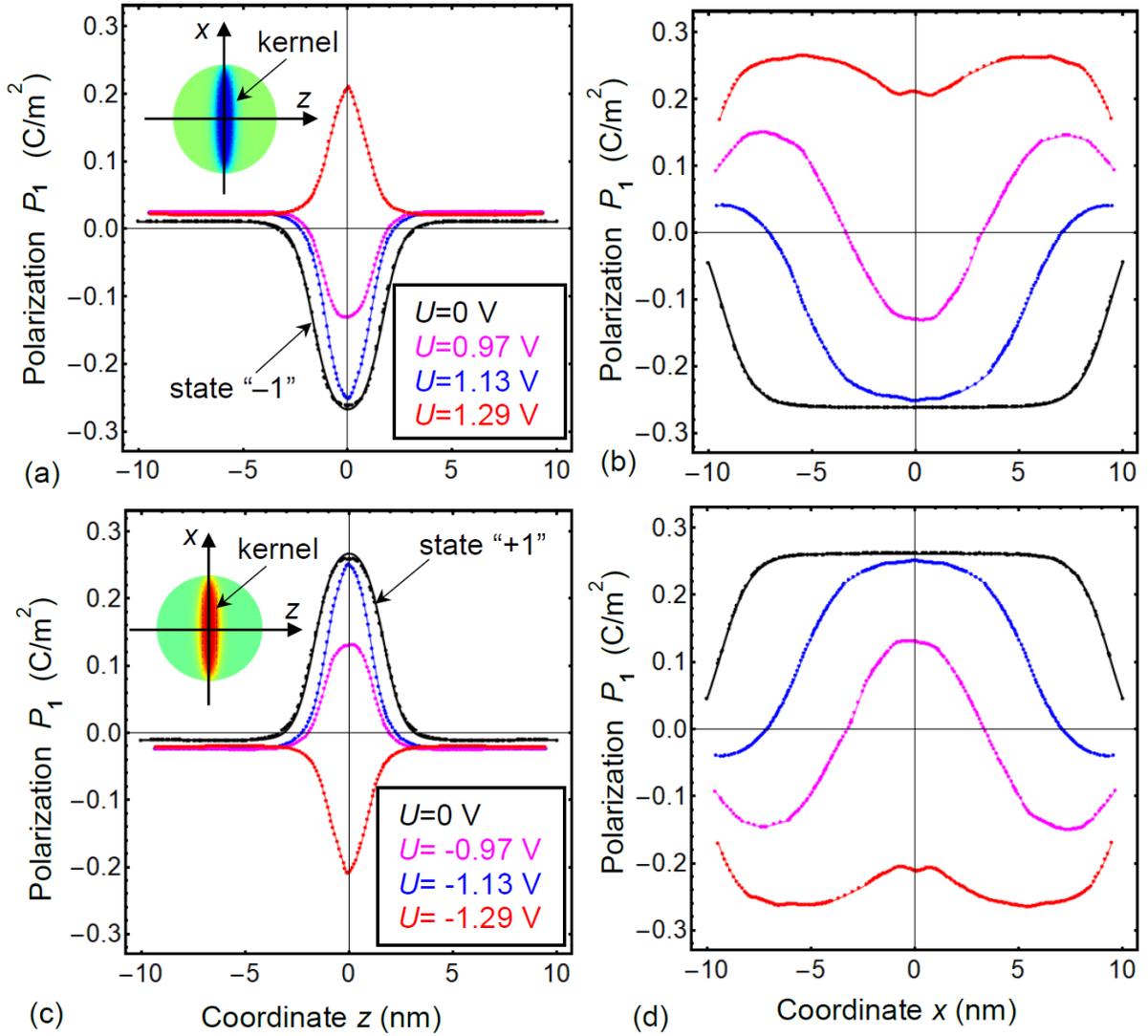

**FIGURE 4.** Polarization profiles calculated at several small voltages $U$ for two perpendicular line scans, $x = y = 0$ **(a, c)** and $y = z = 0$ **(b,d)**. Plots **(a, b)** correspond to $U \geq 0$ [upper part of the voltage pulse in **Fig. 2(b)**], and plots **(c, d)** correspond to $U \leq 0$ [lower part of the voltage pulse in **Fig. 2(b)**]. Symbols are the FEM data. Solid black curves are analytical dependences listed in the text. The voltage cycle period is $10^4 \tau_K$.

We calculated that the average spontaneous polarization of the nanoparticle is $\pm 0.11$ $\mu$C/cm$^2$; and it is pointed along the prolate part of the kernel at $U = 0$ (see color insets in **Fig. 4,** showing "blue" or "red" prolate ellipsoids inside a "green" region). At the same time, one can see the highly polarized kernel with a total dipole moment $\pm 4.11 \times 10^{-26}$ C m and average polarization $\pm 6.58$ $\mu$C/cm$^2$, and the rest of the core with the total dipole moment $\mp 3.66 \times 10^{-26}$ C m and average polarization $\mp 1.03$ $\mu$C/cm$^2$. A much smaller resulting polarization stems from the high compensation degree of the small kernel (15% of the ferroelectric core volume) with high positive (or negative) polarization in the region of small negative (or positive) polarization, whose relative volume is about 85 %.

Note that the switching between the two-fold degenerate ground vortex states "±1" and the excited vortex state "0" can be realized by external electric fields in an ensemble of non-interacting ferroelectric nanoparticles placed in a liquid or polymer matrix.

Our results indicate that the internal kernel structure of three-dimensional vortices is a general feature of nanoscale ferroelectric vortices that has so far not been thoroughly addressed. This is further corroborated by results reported by Mangeri et al. [21] and Pitike et al. [22], who found similar polarization structures in simulations of nanoparticles of other shapes, although the physical origin, energy stability, formation, and in-field and off-field evolution of the ferroelectric vortex with the kernel have not been discussed there. Our findings also correlate nicely with the well-known existence of analogous core structures in ferromagnetic vortices.

### B. Structure of Initial and Final Polarization States

A typical distribution of polarization components and their derivatives before the voltage pulse is initially applied and the final state formed a long time ( $10^4 \tau_K$) after the voltage pulse are shown in **Figs. 5(a)-(d).** The vortex axis coincides with the crystallographic direction [001] prior to the voltage pulse. The direction of external field pulse is [100]. It is seen from **Figs. 5(a)-(b)** that the zero-field (before the pulse) and off-field (relaxed after the pulse) vortex states "0" and "±1" have different axes and internal structures.

The vortex axis coincides with the crystallographic direction [001] prior to the voltage pulse (state "0") [see **Fig. 5(a)**], and its polarization structure is a single curl with a weak fourfold polar anisotropy manifesting itself as the 4 lighter and 4 darker lobes corresponding to the azimuthal angles $\psi = m\pi/8 + n\pi/2$ ($n$ and $m$ are integers). The polar anisotropy is inherent to a tetragonal 4mm phase of a bulk BaTiO$_3$ at room temperature. The polarization derivatives satisfy the equality $\partial P_1/\partial x = -\partial P_2/\partial y$, such that its total divergence div **P** is zero [**Fig. 5(c)**]. Since div **P**=0 in the state "0", the resulting bound charge is absent at the spherical surface in accordance with Gauss' theorem, $\sigma = \int_S (\boldsymbol{P}d\boldsymbol{s}) = \int_V \text{div}\boldsymbol{P}d^3r = 0$. Next, we checked that the resulting dipolar moment $\int_V \boldsymbol{P}d^3r$ is also zero in the state "0".

The domain structure of a ferroelectric core and polarization relaxation in the states "±1" have some distinct features [see **Fig. 5(b)**]. In addition to the vortex part of the polarization itself lying in the plane of the vortex ("simple" vortex), there is a polarization outside this plane localized in an ellipsoidal region elongated along an axis perpendicular to the plane of the vortex. The region looks like a 180-degree prolate "axial nanodomain" directed along the vortex axis [001]. The azimuthal vortex polarization in this nanodomain is absent. It can be seen from **Fig. 5(b)** that the nanodomain has a dipole moment. The dipolar character of the kernel is further evidenced by the potential difference at the opposite ends of the kernel, displayed in the panel on the upper right of **Fig. 2**. The external dipole moment of the kernel is reduced by a partial alignment of the polarization in the nanoparticle along the field of this dipole in regions outside the kernel. This dipole moment of the vortex kernel and its coupling to the azimuthal polarization vortex is an important feature that allows for the manipulation of the vortex axis with a homogeneous external field. There are sub-surface regions near the poles of the axial kernel with nonzero div **P**, which are exactly the oppositely charged regions of the aforementioned dipole [see **Fig. 5(d)**]. It is seen from **Figs. 2** that the vortex states "±1" with a dipolar kernel have a non-zero dipolar moment.

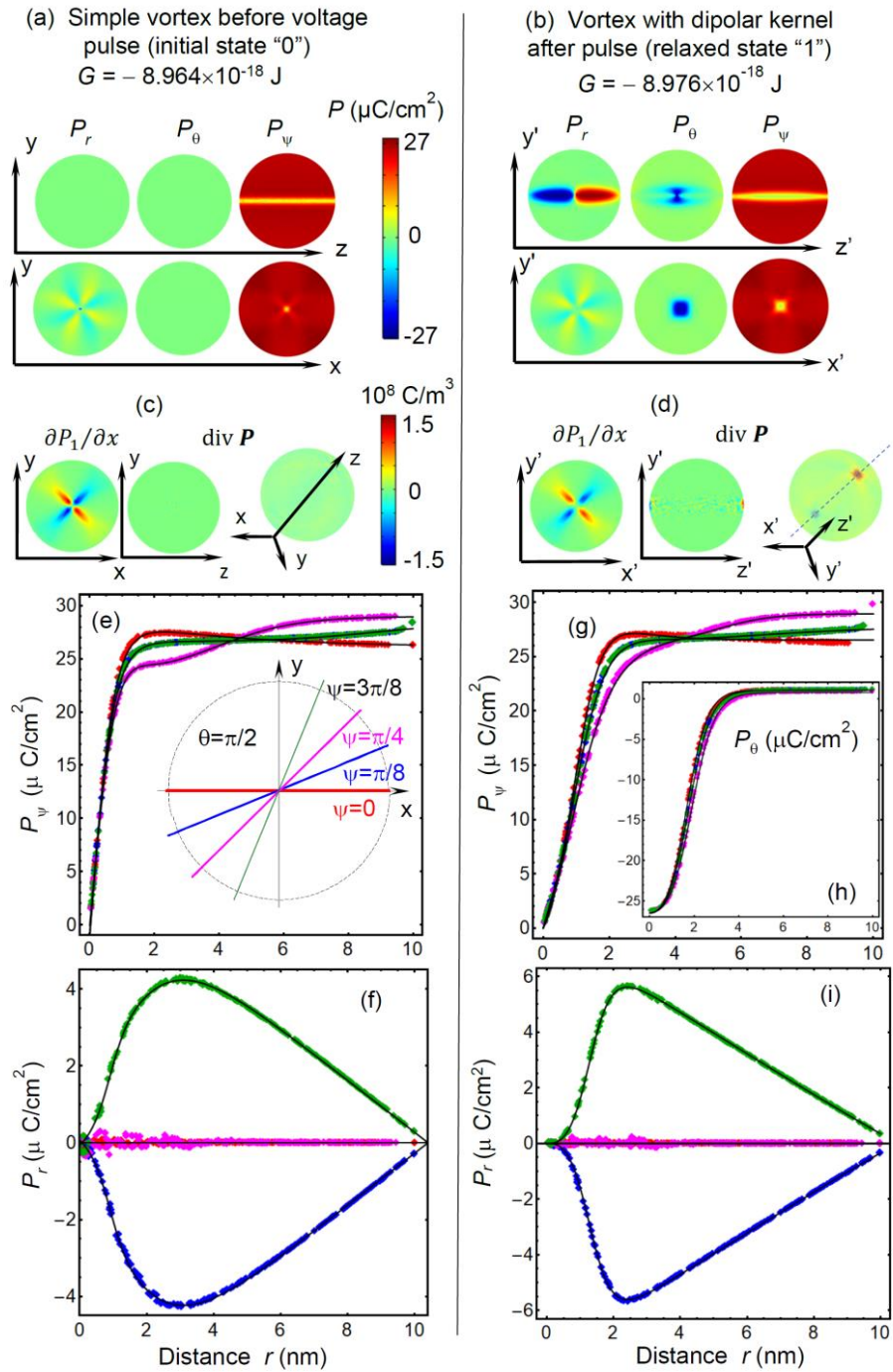

**FIGURE 5.** A typical distribution of polarization components **(a, b)** and their derivatives **(c, d)** calculated by FEM before the voltage pulse application **(a, c)**, and $10^4 \tau_K$ after the pulse **(b, d)**. The vortex axis coincides with the crystallographic direction [001] before the voltage pulse; while its axis coincides with the [100] direction after the pulse and polarization relaxation. The direction of external field pulse is [100]. The radial dependences of the polarization components in spherical coordinates are shown before the pulse in plots **(e-f)**, and after the pulse and system relaxation in plots **(g-i)**. Angle $\theta = \pi/2$ for plots **(e)-(i)**, and the symbols of different colors are calculated by FEM for the angles $\psi = n\pi/2$ (red), $\pi/8 + n\pi/2$ (blue), $\pi/4 + n\pi/2$ (magenta), and $3\pi/8 + n\pi/2$ (green). The solid black curves are calculated from Eqs.(2). Other parameters are the same as in **Fig. 2**.

Allowing for the spherical geometry of the nanoparticle, it appears convenient to calculate and analyze the polarization components in spherical coordinates $\{r, \theta, \psi\}$. The radial distributions of the vortex polarization components before the pulse are shown in **Figs. 5(e-f)**, and the components after the pulse removal and subsequent relaxation are shown in **Figs. 5(g-i)**. The polar angle $\theta = \pi/2$ for plots **(e)-(i)**, and the symbols of different colors, calculated by FEM, correspond to characteristic azimuthal angles $\psi = n\pi/2$ (red), $\pi/8 + n\pi/2$ (blue), $\pi/4 + n\pi/2$ (magenta), and $3\pi/8 + n\pi/2$ (green).

To fit the FEM dependencies shown in **Fig. 5(e)-(i)** let us consider the distribution of the polarization describing the vortex structure in spherical coordinates,

$$\boldsymbol{P} = \{P_r, P_\theta, P_\psi\}. \tag{1}$$

Using the argumentation presented in **Appendix B**, we have "guessed" the following trial functions:

$$P_\alpha(r, \psi, \theta) = \sum_{i=1}^{3} a_i(\psi, \theta) \tanh\left(\frac{r - b_i(\psi, \theta)}{R_i(\psi, \theta)}\right), \qquad \alpha = r, \psi, \theta. \tag{2}$$

For the initial state "0", that is a single vortex without a kernel, the fitting parameters are listed in **Table BI**, **Appendix B**. Solid curves in **Figs. 5(e)-(f)** are calculated analytically from Eq. (2) using these parameters. For a final state "±1", that is a vortex with an axial kernel, the fitting parameters are listed in **Table BII**, **Appendix B**. The solid curves in **Figs. 5(g)-(i)** are calculated analytically from Eq.(2). Note that the polar component $P_\theta(r, \psi, \theta)$ is nonzero for the vortex with an axial kernel. The radial component $P_r(r, \psi, \theta)$ acquires zero values for several angles $\psi = n\pi/4$ defined by ferroelectric anisotropy for both vortex states "0" and "±1".

Since the solid curves in **Figs. 5(e)-(i)** are in a good agreement with the data (symbols) calculated by FEM, we can conclude that the trial functions (2) adequately describe the polarization profiles. Let us analyze the functional form (2), and discuss the meaning of the fitting parameters. The functions have a rather simple physical interpretation of kink-type profiles, inherent to the diffuse Bloch-Ising type domain walls, which are typical for multiaxial ferroelectrics [60]. Fitting parameters $a_i(\psi, \theta)$ and $b_i(\psi, \theta)$ are angle-dependent amplitudes and shifts, respectively. The fitting parameters $R_i(\psi, \theta)$ describe the domain wall width and depend on the wall orientation. Note that $P_r \ll P_\psi$ as anticipated for a single vortex-like state in an anisotropic ferroelectric, since for an isotropic vortex $P_r = P_\theta = 0$ and $P_\psi$ depends only on the distance $r$ from the particle center.

## C. Three-dimensional vortices and Bloch Point Structures: Comparison with Ferromagnets

Analytical expressions (2) are interpolation trial functions, which are selected from general considerations, but not derived. Unfortunately, we are not aware of any analytical model for the structure and dynamics of a ferroelectric or ferromagnetic vortex in nanospheres. At the same time, micromagnetic simulations of vortex structures in ferromagnetic nanodisks [57, 61], mesoscopic islands [62], prisms [63], ellipsoids [64], and magnetic thin films [65] have been previously performed under an external field. These magnetic vortices result from the tendency of the magnetization to form flux-closure patterns, which are frequently found in confined mesoscopic ferromagnets. A typical vortex structure is characterized by the circulation of the in-plane magnetization around a nanometer-sized kernel, whose radius is determined by the competition between the magnetostatic and exchange energies. The functional form of the vortex kernel profile is analyzed in Ref. [66] for different models. At the vortex center, the magnetization rotates out of plane as a result of the exchange interaction, forming an extremely stable structure [67]. The simplest example of a vortex structure occurs in cylindrical-shaped nanomagnets [57, 61] above the single-domain limit [68].

Although the situation in a nanosphere is quite different from that in a cylinder or thin film, it seems possible to extend the above conclusions made for magnetic vortices to ferroelectric nanospheres and underline some important differences between ferromagnetic and ferroelectric vortices. The first important difference is that in contrast to the ferroelectric vortices discussed above, the kernel of ferromagnetic vortices is an integral part of the vortex structure, whose existence is enforced by topology. In a ferromagnetic vortex, the kernel is a consequence of both the smoothness of the magnetization field $\boldsymbol{M}$ in ferromagnets, which is due to the dominant ferromagnetic exchange on short length scales, and the nonlinear constraint $|\boldsymbol{M}| = \text{const.}$, which means that $\boldsymbol{M}$ is a directional field of constant magnitude. In contrast to this, the polarizability of ferroelectric materials makes it possible to have low-dimensional regions representing topological defects, such as Ising-type domain walls where the polarization vanishes on the central plane between two oppositely polarized domains. The only case where such topological defects may occur in homogeneous ferromagnetic materials is the specific three-dimensional structures known as Bloch point structures (**BPS**), which are typically formed at the center of special types of magnetic vortices. For these structures, the transition between oppositely magnetized regions cannot proceed continuously [69, 70], leading to a point with vanishing magnetization, **M**=0. The simplest topological configuration of the Bloch point has the form $\mathbf{M} = \mp M\boldsymbol{e_r}$.

For a ferroelectric vortex state "0" we calculated that $\mathbf{P} = P\boldsymbol{e_r}$ around its axis $z$ [see Eqs.(2) and **Fig. 5(f)** at $U = 0$]. Thus, formally we can interpret the line $x = y = 0$ in this state as an "Ising line" inside the vortex [71], because the "line" is an analogue of a diffuse Bloch-Ising type domain wall inherent to ferroelectrics [60]. BPSs appear when a vortex interacts with an external field and transforms into the state "±1" with a field-induced kernel. Typical BPSs with two diametrically opposite Bloch-points ($\mathbf{P} = 0$) located at the core surface are shown in **Fig. 6** for the voltage $U = -1.38$ V.

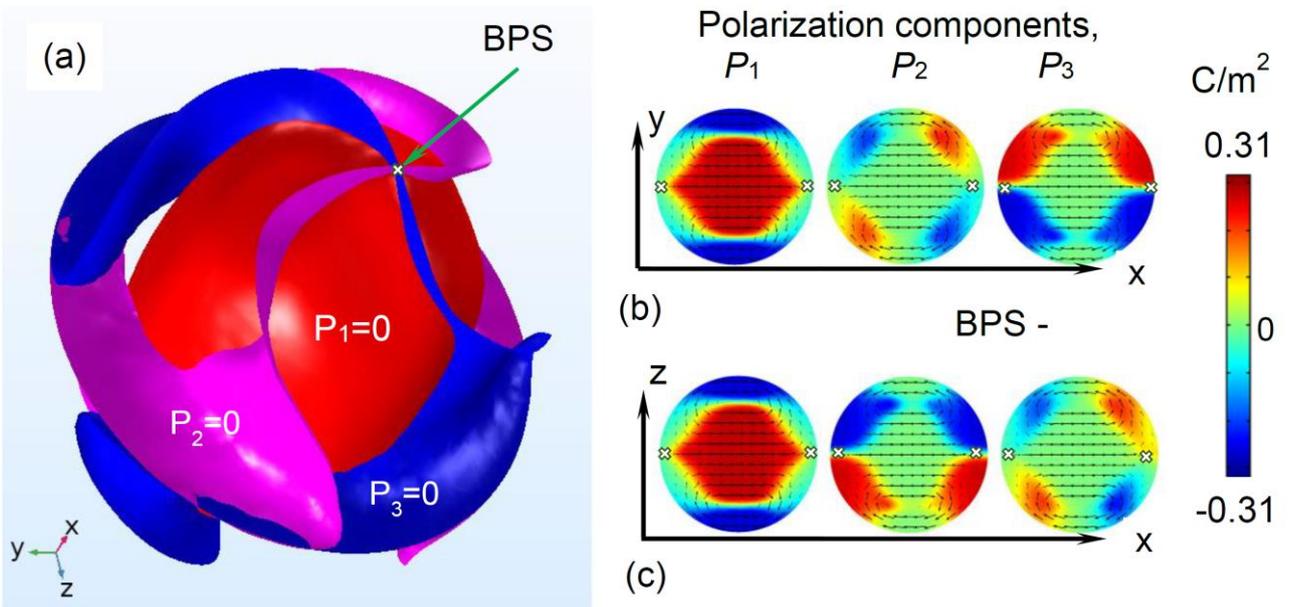

**FIGURE 6.** Determination of a Bloch point structure (BPS) position. **(a)** The intersection points (denoted with a white cross) of the polarization components' iso-surfaces, $P_1 = 0$ (red), $P_2 = 0$ (magenta), and $P_3 = 0$ (blue), show the position of two diametrically opposite Bloch-points ($\mathbf{P}$=0) located near the core surface. Cross-sections z = 0 **(b)** and y = 0 **(c)** of the nanoparticle polarization with Bloch points (denoted with a cross) for three components of the polarization, $P_1$ (left column), $P_2$ (middle column), and $P_3$ (right column). Parameters: voltage amplitude $U = -1.38$ V and period $10^3\tau_K$. Small black arrows indicate the direction of polarization in plots (b)-(c). Other parameters are the same as in **Fig. 2(a).**

To the best of our knowledge, the BPS are new topological structures in ferroelectrics. Remarkably, our results evidence a close analogy with micromagnetism concerning the BPS, because there, too, they typically appear during switching processes. In magnetism, this was discovered more than 40 years ago by Arrott et al. [72], where the BPS were called "point singularities". It is worth noting that the Bloch points propagate from both sides of the sample during a ferromagnetic or ferroelectric switching process. This results in the temporary formation of an isolated, non-reversed region inside the ferromagnet called a "drop" by Hertel and Kirschner [73], or in the formation of the ferroelectric kernel in the considered case. These observations allow us to conclude that there are important similarities between magnetic BPS and the polarization BPS that we calculated in the ferroelectric spherical nanoparticle.

### D. Polarization Response to a Periodic Electric Field

The dependencies of the polarization **P** averaged over the core volume, with a periodic voltage $U \sin\left(\frac{2\pi t}{T_p}\right)$ applied between the electrodes, were studied for different periods $T_p$ and directions of the nanoparticle core crystallographic axes. The homogeneous electric field $\mathbf{E}^{ext}$ is perpendicular to the electrodes [see **Fig. 1(b)**]. The FEM simulation showed that a small dipolar kernel surrounded by a curling part of the polarization can be formed at rather small voltages (~15 mV) during the first period of the electric field application. The induced polarization is parallel to the field, $\mathbf{P} \uparrow\uparrow \mathbf{E}^{ext}$, and increases with an increase of $\mathbf{E}^{ext}$. The first cycle with a transient process of the initial vortex destruction or reorientation is not shown in the **Figs. 6-8**.

The quasi-static dependence $P(U)$, shown in **Fig. 7**, was calculated for $T_p = 10^4 \tau_K$ and $\mathbf{E}^{ext}$ parallel to one of the crystallographic directions [100], [010], or [001]. The initial vortex axis coincides with the direction [001], and the angle between the direction [001] and $\mathbf{E}^{ext}$ influences the polarization during the first cycle of electric field. After the first cycle, $P(U)$ becomes the same for $\mathbf{E}^{ext}$ parallel to the direction [100], [010], or [001].

The quasi-static behavior of the polarization dependence $P(U)$, shown in **Fig. 7**, can be explained as follows. The spontaneous polarization is conditioned by the small average polarization of the kernel, $P(0) = \pm P_k$ (see zoomed inset to **Fig. 7**). The value of the field-induced polarization $P(U)$ is linearly proportional to the applied voltage at $|U| < U_{cr}$, where the critical voltage $U_{cr} \approx$ 7 V, and the internal structure of the vortex polarization continuously changes at $|U| < U_{cr}$. To illustrate the effect, color images show the polarization distribution at positive and negative voltages $U$. The images' location corresponds to the counterclockwise direction along the loop.

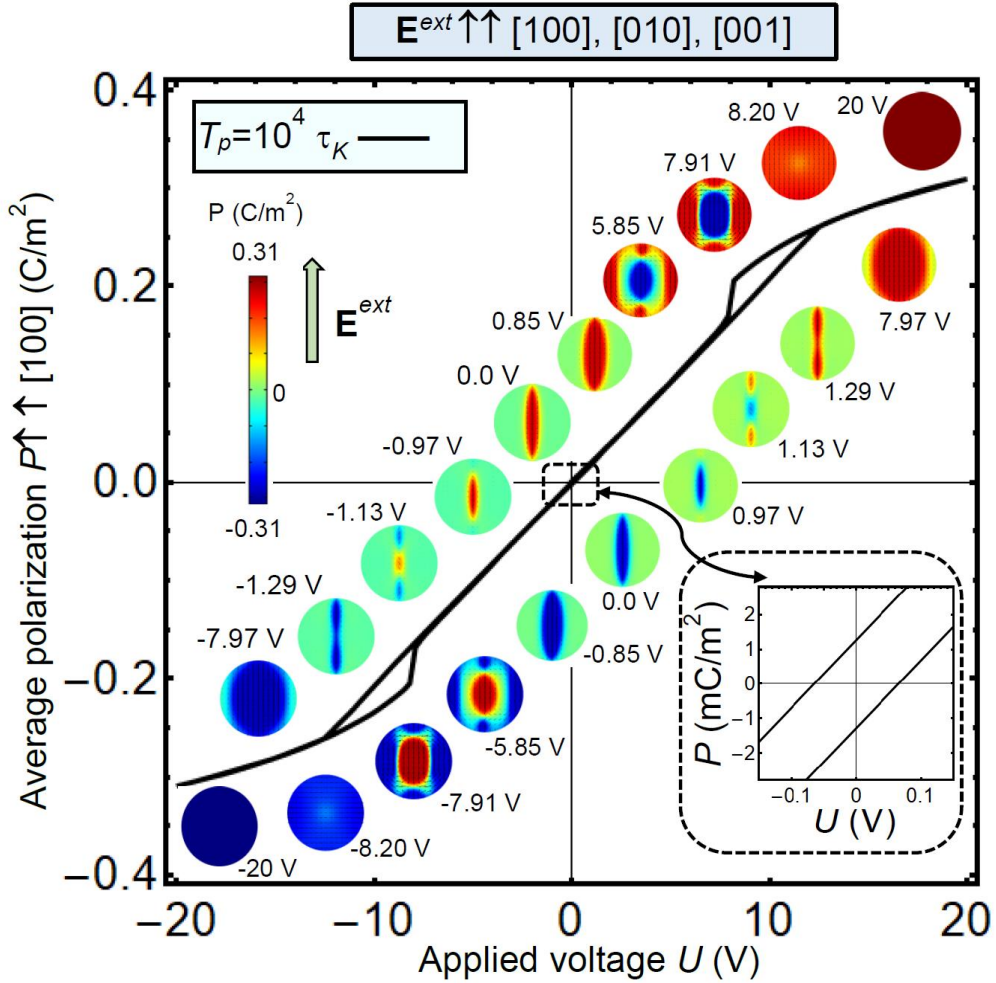

**FIGURE 7.** Quasi-static hysteresis loop of the average polarization $P(U)$ for several directions of the external field $\mathbf{E}^{ext} \uparrow\uparrow [100]$, [101], and [001]. The spontaneous polarization conditioned by the dipolar kernel is shown in the inset. Small color images show polarization distributions at different voltages $U$; the values are indicated next to each image. Polarization cross-sections, shown in the images, are parallel to $\mathbf{E}^{ext}$. Small black arrows indicate the direction of polarization in color images. The period of the applied voltage $T_p = 10^4 \tau_K$ and other parameters are the same as in **Fig. 2(b)**.

Since $P(U) = \pm P_k + AU$ at $|U| < U_{cr}$, the average polarization is linearly proportional to the effective electric field $E_{eff}(U)$ inside the particle core:

$$P(U) = \pm P_k + \varepsilon_0 \chi_{eff} E_{eff}(U), \tag{3a}$$

where $\pm P_k$ is the small spontaneous polarization of the kernel, $\varepsilon_0$ is the universal dielectric constant, and $\chi_{eff}$ is the effective linear dielectric susceptibility of the ferroelectric core. The kernel polarization is bi-stable, because it can be positive or negative depending on the field gradient (increase or decrease). The effective field $E_{eff}(U)$ can be estimated from the expression,

$$E_{eff}(U) = 9\varepsilon_e \varepsilon_s \eta(\varepsilon_f) E^{ext}(U), \qquad \eta(\varepsilon) = \frac{(1+\Delta)^3}{2(\varepsilon_e - \varepsilon_s)(\varepsilon_s - \varepsilon) + (1+\Delta)^3 (2\varepsilon_e + \varepsilon_s)(\varepsilon + 2\varepsilon_s)}. \tag{3b}$$

Thus $E_{eff}$ is proportional to an external field $E^{ext}(U)$ multiplied by a depolarization factor $\eta(\varepsilon_f)$. The factor depends on the relative thickness of the shell, $\Delta = \frac{\Delta R}{R}$; $\varepsilon_f$, $\varepsilon_S$, and $\varepsilon_e$ are the ferroelectric core, paraelectric shell, and surrounding media permittivities, respectively. Equation (3b) is derived in **Appendix C** (section **C.2.1**).

The value $\chi_{eff} \approx 360$ was estimated from the curve slope in **Fig. 7** using the relationship $E^{ext} \approx \frac{U}{2(R+\Delta R)}$. Note, the value is essentially higher than the relative susceptibility $\chi_{11}^{BTO} \approx (110-170)$ of a bulk BaTiO$_3$ in the polar direction [001], and much smaller than $\chi_{11}^{BTO} \approx 4000$ in the perpendicular direction [100] at 298 K. The value $\chi_{eff}$, being surprisingly close to the permittivity $\varepsilon_s \approx 300$ of a paraelectric SrTiO$_3$ shell, can indicate the appearance of super-paraelectric features in small core-shell ferroelectric nanoparticles.

As the voltage overcomes the critical value of the vortex destruction, the curling part of the polarization disappears and the ferroelectric core becomes single-domain. In the considered case, $U_{cr1} = 7$ V corresponds to the first quasi-static critical field along the main crystallographic directions. At $U > U_{cr1}$ the core remains single-domain up to the maximal voltage (20 V in the considered case). Next, when the voltage is reduced to the next critical value $U_{cr2} = 12$ V corresponding to the second quasi-static critical field along the direction [001], the polarization becomes curled again, but the vortex does not appear immediately. At first the domains with polarization modulated in the plane perpendicular to the field axis appear. These two different critical fields determine the quasi-static double hysteresis loop $P(U)$ shown in **Fig. 7**.

The dynamics of the average polarization, $P(U)$, shown in **Fig. 8**, is calculated for **E**$^{ext}$ parallel to: one of the crystallographic direction [100], [010], or [001] [see plot **(a)**], one of the face diagonal [101], [101] or [101] [see plot **(b)**], or the main diagonal [111] of the cubic cell [see plot **(c)**]. The initial vortex axis coincides with the crystallographic direction [001].

Dash-dotted green, black solid, red dashed, and blue dotted curves correspond to the relatively long ($T_p = 10^4\tau_K$), intermediate ($T_p = 10^3\tau_K$), short ($T_p = 30\tau_K$), and ultra-short ($T_p = 3\tau_K$) periods of the applied voltage, respectively. At $U = 0$, the case of the fast cycling (blue dotted and red dashed curves) shows evident remanence, while the slower cycling (dash-dotted green and black solid lines) corresponds to much smaller remanence, visible only in the inset to **Fig. 7**. A non-zero remanence is an inherent feature of the vortex kernel. We attribute the reduction of the remanence at slower cycles to a comparatively slow relaxation of the larger vortex region surrounding the kernel, such as to partially align the vortex polarization with the depolarization field of the kernel, i.e. in a direction opposite to the kernel polarization. Actually, the color images in **Fig. 8(a)**, which show polarization distributions calculated for $T_p = 10^3\tau_K$ at different voltages $U$, are very different from the images in **Fig. 7**, calculated for $T_p = 10^4\tau_K$.

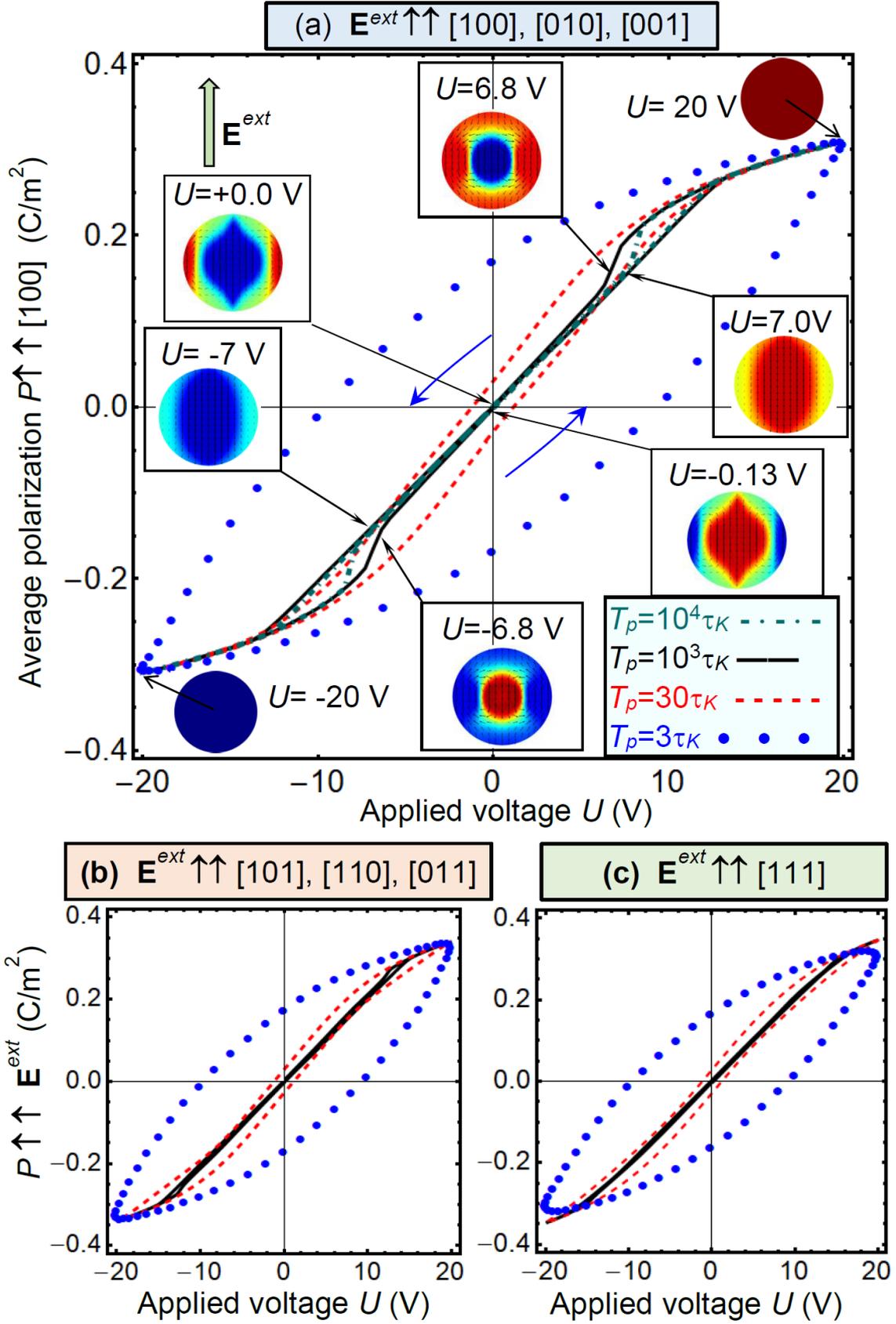

**FIGURE 8.** Hysteresis loops of the average polarization $P(U)$ calculated for several directions of an external field $\mathbf{E}^{ext} \uparrow\uparrow [100]$ **(a)**, $\mathbf{E}^{ext} \uparrow\uparrow [101]$ **(b)**, and $\mathbf{E}^{ext} \uparrow\uparrow [111]$ **(c)**. Dash-dotted green, black solid, red dashed, and blue dotted curves correspond to different periods $T_p = (10^4 - 3)\tau_K$ of the applied voltage $U$. Color images in plot (a) show polarization distributions calculated for $T_p = 10^3 \tau_K$ at different voltages $U$, which are indicated next to each image. Small black arrows indicate the direction of polarization in the color images. Other parameters are the same as in **Fig. 2(a)**.

It is seen that the polarization $P(U)$ induced by the electric field with a period $T_p = (10^4 - 10^3)\tau_K$ is quasi-linear in the region $|U| < 5V$ [see green dash-dotted and black solid curves in **Fig. 8(a)**]. The linear part is followed by either a change of the slope for $\mathbf{E}^{ext} \uparrow\uparrow [111]$, or by the appearance of rather narrow minor loops at $5V < |U| < 14V$ for $\mathbf{E}^{ext} \uparrow\uparrow [100]$ [compare **Fig. 8(a)-(c)**]. The sub-linear increase of $P(U)$ starts at higher voltages, being the most pronounced for $\mathbf{E}^{ext} \uparrow\uparrow [100]$. A rather slim hysteresis loop $P(U)$ opens with decreasing $T_p$; and these loops are almost the same for different directions of $\mathbf{E}^{ext}$ (see red dashed curves calculated for $T_p = 30\ \tau_K$). The loops acquire a quasi-elliptic shape for a small period $T_p = 3\tau_K$ (see blue dotted curves calculated for $T_p = 3\ \tau_K$).

## IV. ELECTROSTATIC FORCES ACTING ON THE CORE-SHELL NANOPARTICLE

Using FEM, we calculated the electrostatic forces acting on the core-shell nanoparticle placed in a liquid (or viscous) medium under an external electric field. The electric field $\mathbf{E}$ in the ferroelectric core consists of external and depolarization contributions, $\mathbf{E} = \mathbf{E}^{ext} + \mathbf{E}^d$, where $\mathbf{E}^d$ is generally inhomogeneous. The only exceptions are homogeneously polarized ellipsoids, for which the internal depolarization field is homogeneous [74]. Since free charges are absent inside the core, $\text{div}\mathbf{D} = 0$ and $\mathbf{D} = \mathbf{P} + \varepsilon_0 \mathbf{E}$. The electric displacement is continuous at the core-shell interface, but the radial component of the electric field and polarization are discontinuous at the interface due the absence of ferroelectric polarization $\mathbf{P}_S$ in the shell and the different dielectric permittivities, $\varepsilon_f$ and $\varepsilon_s$.

The electrostatic force $\mathbf{F}$ and torque $\mathbf{M}$ components acting on a core-shell nanoparticle placed in the external electric field $\mathbf{E}^{ext}$ can be calculated from the expressions [74]:

$$\mathbf{F} = \int_{V_c} (\mathbf{P}_c(\mathbf{r}') \cdot \nabla') \mathbf{E}^{ext}(\mathbf{r}') dV' + \int_{V_S} (\mathbf{P}_s(\mathbf{r}') \cdot \nabla') \mathbf{E}^{ext}(\mathbf{r}') dV' , \qquad (4a)$$

$$\mathbf{M} = \int_V ([\mathbf{P}(\mathbf{r}') \times \mathbf{E}^{ext}(\mathbf{r}')] + [\mathbf{r}' \times (\mathbf{P}(\mathbf{r}') \cdot \nabla') \mathbf{E}^{ext}(\mathbf{r}')]) dV' , \qquad (4b)$$

where summation on $i = 1, 2, 3$ is performed, the polarization $\mathbf{P}_c = \mathbf{P}_S + \varepsilon_0 (\varepsilon_f - 1)\mathbf{E}$ in the ferroelectric core, and $\mathbf{P}_{sh} = \varepsilon_0 (\varepsilon_s - 1)\mathbf{E}$ in the paraelectric shell. The scalar product $(\mathbf{P} \cdot \nabla) = P_i \frac{\partial}{\partial x_i}$, and the integration is performed over the core and shell regions with total volume $V = V_c + V_s$. Following Landau et al. [74], $\mathbf{E}^{ext}$ is calculated under the absence of $\mathbf{P}$.

## A. Electrostatic Force and Torque Acting on a Nanoparticle Placed in a Homogeneous Electric Field

As it follows from Eq.(4a) the force is zero ($F = 0$) for any constant field, because $P_i \frac{\partial E_i^{ext}}{\partial x_i} = 0$ for $\frac{\partial E_i^{ext}}{\partial x_j} = 0$. The polarization vortex (with or without a dipolar kernel) cannot move translationally in a homogeneous external field; it can only rotate around its axis [see **Fig.1(b)**]. Correspondingly, the torque $\boldsymbol{M}$ in a homogeneous field is $\boldsymbol{M} = -\left[\boldsymbol{E}^{ext} \times \int_V \boldsymbol{P}(r')dV'\right]$. It is easy to check that $\int_V \boldsymbol{P}(\boldsymbol{r}')dV' = 0$ for the vortex state "0" without a kernel, thus $\boldsymbol{M}$=0 in this state.

FEM shows that the average value of the "vortex + kernel" spontaneous polarization $\boldsymbol{P}_k$ is about $\pm 10^{-3}$ C/m$^2$ in the absence of an external field. When the external homogeneous field is present, the average polarization $\boldsymbol{P}$ becomes almost collinear to the field (see e.g. **Figs. 7-8**). This happens rapidly (at times $< 30\tau_K$) via several mechanisms, such as nanoparticle rotation in a liquid medium, polarization re-orientation, and induction by the $\boldsymbol{E}^{ext}$. As a result, the torque acting on the particle is small, except for the time intervals when the external field and nanoparticle axes change their mutual orientation. Assuming that the only non-zero component of the average polarization is $\bar{P}_z$, the transient torque is:

$$\mathbf{M} = \frac{4\pi}{3} R^3 \left(E_y^{ext}\boldsymbol{e}_x - E_x^{ext}\boldsymbol{e}_y\right)\bar{P}_z, \tag{5}$$

where $\bar{P}_z = \frac{1}{V}\int_V P_z(\mathbf{r})dV'$ (see Section **C.1** in **Appendix C** for mathematical details). Evolution of the normalized torque $M_y$ acting on the core-shell nanoparticle is shown in **Fig. 9**. The external field is directed in the crystallographic direction $z$ over a time that is long enough ($>100\ \tau_K$) to orient the nanoparticle polarization $\boldsymbol{P}$ in the z-direction (red curve with label "$E_z$"). The torque is absent for the case $\boldsymbol{P} \uparrow\uparrow \mathbf{E}^{ext}$. At this point the z-directed field is removed and a subsequent field is applied in the x-directed field over a period of one $\tau_K$ (blue curve with label "$E_x$"). As one can see from **Fig. 9**, the maximal value of the torque is achieved with some delay after the electric field in both directions have comparable values. This is due to a certain retardation in the polarization response to the field, by an order of the Landau-Khalatnikov relaxation time $\tau_K$, which can be seen from the distribution of polarization components (see color images in **Fig. 9**). The maximal value of the torque is achieved at the time when the voltages $U_z = 0.05$ V and $U_x = 14.63$ V.

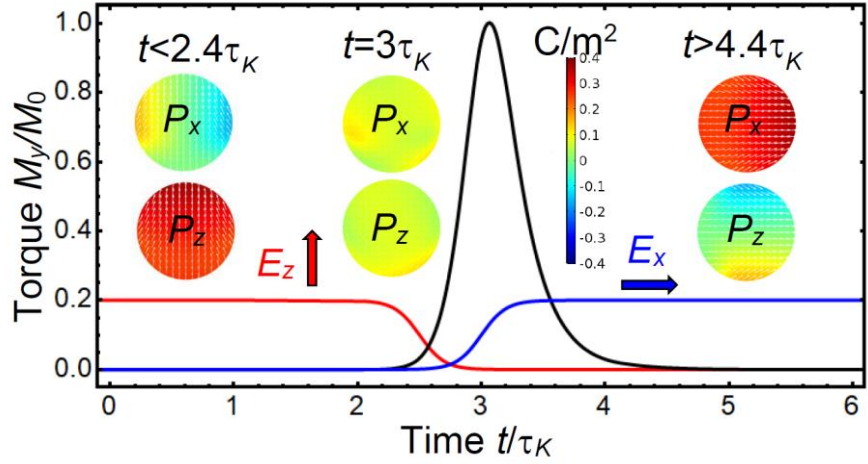

**FIGURE 9.** Normalized torque $M_y$ (black curve) acting on the core-shell nanoparticle vs. time. At first, the field in z-direction ($E_z$, red curve) is applied, then it is turned off and quickly applied in the orthogonal direction ($E_x$, blue curve). Color images show the distribution of polarization components inside the core at several moments of time. Parameters $U_{max}$ =20 V; $h$ = 17 nm, and the factor $M_0 = 0.965\ 10^{-16}$ J. White arrows indicate the direction of polarization. Other parameters are the same as in **Fig. 3.**

From the expression (5) and **Fig. 9** we can conclude that the torque **M** tends to rotate the nanoparticle in such way that the vortex axis ([001] in the considered case) or the polarization vector becomes parallel to the field direction. This conclusion remains valid for an inhomogeneous electric field and is used in the following sections.

## B. Electrostatic Force Acting on a Nanoparticle Placed in a Strong Electric Field Gradient

As it follows from the above discussion, the studied spherical nanoparticles with a vortex polarization do not have any net electrostatic charge (monopole moment), and therefore such particles do not experience any force in a homogeneous electric field. To study the effect of an *inhomogeneous* field, it is necessary to analyze the particle dipole moment. If the nanoparticle core is in the state "0", representing a vortex without a central axial kernel, then it can be recognized immediately that the dipole moment is zero for symmetry reasons. However, if the vortex possesses a kernel forming an axial nanodomain (states "±1"), then the kernel has a sizable dipole moment, and in that case an inhomogeneous external field will exert a force on the nanoparticle. The experimentally observed particle motion in the inhomogeneous electric field created by a charged tip or wire [42-47] can thus be attributed to the dipole moment of the vortex kernel, which (if not already present) might be induced into the vortex structure. The dipole moment of the vortex kernel leads to an attractive gradient force in the inhomogeneous electric field: $\mathbf{F} = (\mathbf{p} \bullet \nabla)\mathbf{E}$ [74]. In this scenario, however, it is assumed that the vortex structure remains intact, and that the inhomogeneous field only generates the nanodomain as the kernel of the vortex without destroying it. Actually, **Figs. 3, 7, & 8** demonstrate that the vortices can be partially or completely destroyed by irrotational external fields, and the particle polarization follows the external field.

As a next step, the electrostatic force acting on a core-shell nanoparticle is calculated. The ferroelectric core has an average spontaneous polarization $\boldsymbol{P}_k = \{P_{k1}, P_{k2}, P_{k3}\}$, which originates from the polarized vortex kernel. The particle is placed in a strongly inhomogeneous electric field that is produced by the charged tip electrode [see **Fig. 1(c)**]. For this case, the effective point charge model [75] is applicable. The effective charge $Q^*$ is located at the point $\boldsymbol{h} = \{0, 0, h\}$, where $h > R + \Delta R$ corresponding to the charge location outside the particle. The effective charge $Q^* \approx C_t U$, where $C_t$ is the tip effective capacity and $U$ is applied voltage. For a spherical tip $C_t = 4\pi\varepsilon_0\varepsilon_e R_t$, where $R_t \sim (5-50)$ nm is the tip curvature.

After lengthy calculations made in the Section **C.2.2** of **Appendix C**, we derived that the force consists of two contributions,

$$\boldsymbol{F} = \boldsymbol{F}_C + \boldsymbol{F}_Q. \tag{6a}$$

The first force component, $\boldsymbol{F}_C$, is proportional to the core spontaneous polarization, i.e. it is proportional to the kernel dipole moment. $\boldsymbol{F}_C$ is expressed via an effective dipole moment, $\boldsymbol{p}_f$:

$$\boldsymbol{F}_C = \frac{Q^*}{4\pi\varepsilon_0\varepsilon_e h^3}\left[\boldsymbol{p}_f - 3(\boldsymbol{p}_f \cdot \boldsymbol{h})\frac{\boldsymbol{h}}{h^2}\right], \qquad \boldsymbol{p}_f = 12\pi\varepsilon_s\varepsilon_e\eta(\varepsilon_f)R^3\boldsymbol{P}_k. \tag{6b}$$

Note that the depolarizing factor $\eta(\varepsilon_f)$ has the same functional form as in Eq.(3b), $\eta(\varepsilon_f) = \frac{(1+\Delta)^3}{2(\varepsilon_e-\varepsilon_s)(\varepsilon_s-\varepsilon_f)+(1+\Delta)^3(2\varepsilon_e+\varepsilon_s)(\varepsilon_f+2\varepsilon_s)}$, and depends on $\varepsilon_f$, which is the sum of background permittivity $\varepsilon_b$ and effective susceptibility $\chi_{eff}$, $\varepsilon_f = \varepsilon_b + \chi_{eff}$. In accordance with Eqs.(3), the estimate $\varepsilon_f \approx 365$ is valid for voltages $|U| < U_{cr}$. The force (6b) is a typical dipole force decreasing with distance as $1/h^3$.

The second contribution, $\boldsymbol{F}_Q$, is a dielectrophoretic force [76],

$$\boldsymbol{F}_Q = \frac{-Q^{*2}}{4\pi\varepsilon_0}\left(\sum_{n=1}^{\infty}\frac{R^{1+2n}}{h^{3+2n}}G_n(\varepsilon_f)\right)\frac{\boldsymbol{h}}{h}, \tag{6c}$$

with an expansion coefficient

$$G_n(\varepsilon) = \frac{n(1+n)(1+\Delta)^{1+2n}\left[(1+\Delta)^{1+2n}(\varepsilon_e-\varepsilon_s)(n\varepsilon+(1+n)\varepsilon_s)+(\varepsilon_s-\varepsilon)(n\varepsilon_e+(1+n)\varepsilon_s)\right]}{n(1+n)(\varepsilon-\varepsilon_s)(\varepsilon_s-\varepsilon_e)+(1+\Delta)^{1+2n}(\varepsilon_e(1+n)+n\varepsilon_s)(n\varepsilon+(1+n)\varepsilon_s)}. \tag{6d}$$

The dielectrophoretic force decreases with distance as $1/h^5$.

If the particle is far enough from the probe tip ($h \gg R + \Delta R$), only the first term with $n=1$ is significant in Eq.(6c), and the force components acquire a much simpler form:

$$F_1 \approx 3\varepsilon_s\eta(\varepsilon_f)\frac{Q^*}{\varepsilon_0}P_{k1}\left(\frac{R}{h}\right)^3, \qquad F_2 \approx 3\varepsilon_s\eta(\varepsilon_f)\frac{Q^*}{\varepsilon_0}P_{k2}\left(\frac{R}{h}\right)^3, \tag{7a}$$

$$F_3 \approx -\left[6\varepsilon_s\frac{Q^*}{\varepsilon_0}P_{k3} + \frac{Q^2}{2\pi\varepsilon_0 h^2}\mu(\varepsilon_f)\right]\eta(\varepsilon_f)\left(\frac{R}{h}\right)^3, \tag{7b}$$

where the dielectric factor $\mu(\varepsilon_f) = (1 + \Delta)^3(\varepsilon_e - \varepsilon_s)(\varepsilon_f + 2\varepsilon_s) + (\varepsilon_s - \varepsilon_f)(\varepsilon_e + 2\varepsilon_s)$ is introduced. If the nanoparticle is far enough from the tip the force component $F_3$ scales as $A\left(\frac{R}{h}\right)^3 + B\left(\frac{R}{h}\right)^5$; thus the dipole force $\sim A\left(\frac{R}{h}\right)^3$ dominates with increasing distance.

The case $\boldsymbol{P}_k = \{0, 0, \pm P_k\}$ corresponding to an already rotated nanoparticle with two orientations of the kernel polarization is considered below. For this case, the only nonzero component is $F_3$. The total electrostatic force $F$ acting on the core-shell nanoparticle is $F = F_Q + F_C$ [see Eq.(6a)]; the total force $F$ and its contributions, $F_C$ and $F_Q$, are shown in **Fig. 10.** Note that $F_C$ can change its sign depending on the sign of $P_k$, namely $F_C > 0$ for $Q^*P_k > 0$ and $F_C < 0$ for $Q^*P_k < 0$. Two cases, $F = F_Q + |F_C|$ and $F = F_Q - |F_C|$, are shown by black and magenta curves, respectively. The total force can go to zero when $F_C$ and $F_Q$ counteract each other, it can even change direction when $|F_C| > F_Q$. **Figure 10(a)** illustrates the dependence of the force on the voltage $U$ calculated at a fixed distance, $h = 500$ nm. The dependence of the force on the distance $h$ calculated at a fixed voltage $U = 100$ mV is shown in **Fig. 10(b)**.

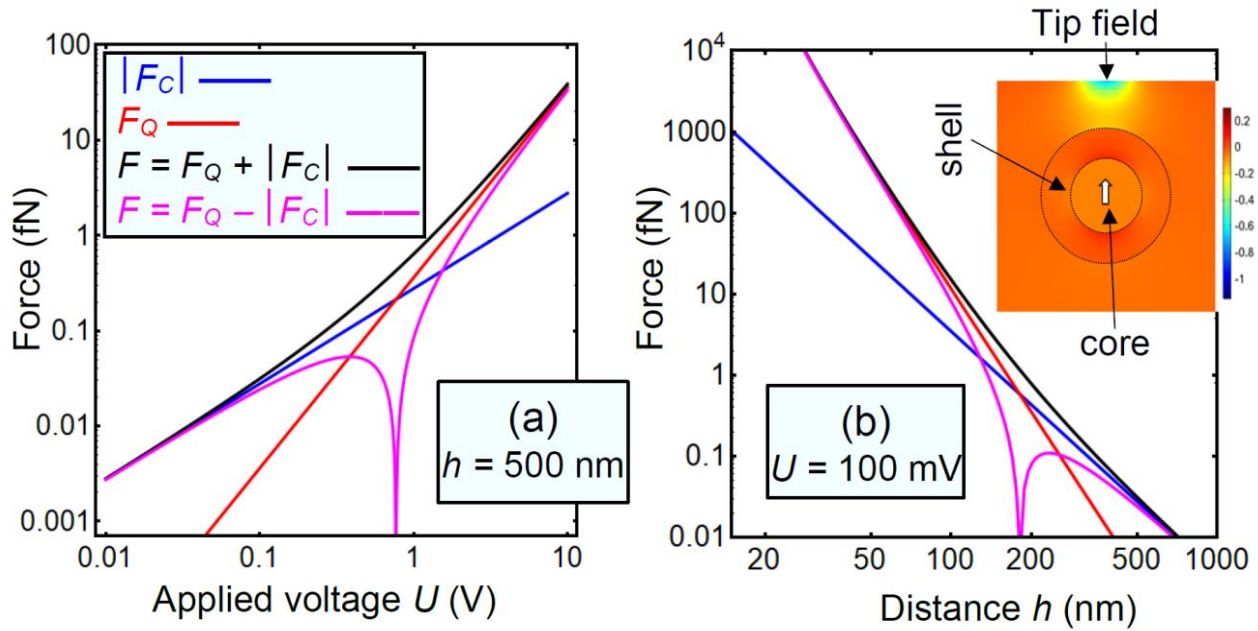

**FIGURE 10. (a)** Electrostatic force $F$ (black and magenta curves) and its contributions, the absolute value of $F_C$ (blue curves) and $F_Q$ (red curves), acting on the core-shell nanoparticle vs. the voltage $U$ applied to the electrode tip located at distance $h = 500$ nm. **(b)** $F$, $F_C$, and $F_Q$ vs. the distance $h$ at $U = 100$ mV. Curves are calculated from Eqs.(6) for a nanoparticle spontaneous polarization $P_k = \pm 0.11$ μC/cm², radius $R = 10$ nm, shell thickness $\Delta R = 4$ nm, $\varepsilon_e = 10$, $\varepsilon_s = 300$, $\varepsilon_f = 365$, and tip capacity $C_t = 5\cdot10^{-17}$ F corresponding to the curvature $R_0 = 50$ nm. The force scale is $10^{-15}$ N (femtoNewtons, fN).

# V. POSSIBLE APPLICATIONS IN LOGIC UNITS

Since the vortex axis should have significantly enhanced electro-conductivity in comparison to the rest of the particle volume [77, 78], it can act as a conductive nanosized channel. Actually, as one can see from **Fig. 11**, the bound charge $\sigma$ being equal to $P_r(R)$, is maximum at the contact point of the vortex axis to the surface. Note that pronounced red ($\sigma > 0$) and blue ($\sigma < 0$) maxima of the bound charge exist where the prolate dipolar kernel of the vortex contacts the surface of the core; they are absent for a simple vortex [compare **Figs. 11(a)-(b)** with **Figs. 11(c)-(d)**]. Such surface charges of opposite sign on opposite sides are characteristic for a dipole. In addition to this primary axial dipole moment, we also observe an anisotropy-induced azimuthal modulation leading to higher-order moments that represent a much weaker effect.

Free charges from the surrounding medium screen the bound charges, and in turn significantly enhance the surface conductivity around the vortex axis. The enhanced conductivity of the nanodomain walls, which is almost uncharged inside the BaTiO$_3$ core, and its physical nature related to flexoelectricity and deformation potential has been studied in detail in Ref.[79]. Note that the vortex axis with an axial kernel should be conductive in other multiaxial ferroelectrics such as Pb$_x$Zr$_{1-x}$TiO$_3$ [80] or BiFeO$_3$ [81, 82]. It should also be noted that a ferroelectric nanoparticle with significantly higher kernel polarization can possibly be fabricated using tunable coatings [83].

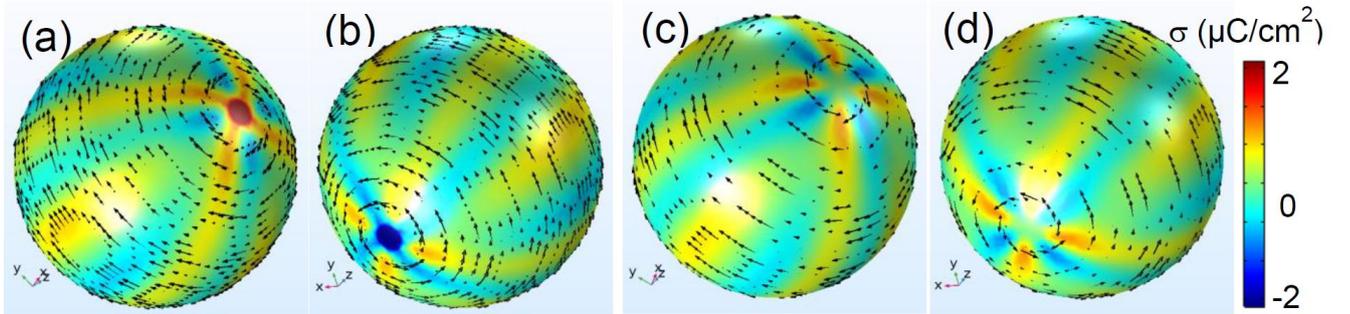

**FIGURE 11.** Bound charge distribution at the surface of a BaTiO$_3$ nanoparticle covered with a SrTiO$_3$ shell. The surface charge of vortex with an axial kernel is shown in plots **(a) − (b)**, and the charge for a simple vortex is shown in plots **(c) − (d)**. Black arrows indicate the direction of polarization. An external field is absent. Other parameters are the same as in **Fig. 2.**

The energy of the vortex state "0" demonstrates a 6-fold degeneracy consisting in the 3 equally probable axis directions multiplied by clockwise and counterclockwise directions of polarization rotation across the vortex axis. For the vortex state "±1" with a kernel, the degeneracy becomes 12-fold, since the above 6-fold degeneracy doubles by 2 equiprobable signs of polarization inside the vortex kernel. The 18-fold degeneracy can be obtained as a sum of 6 vortex states "0" and 12 states "±1" of vortex with an axial kernel. As we have established in this work, the manifold degeneracy of vortex states corresponds to three energy minima, where an "excited" vortex state "0" and the vortex ground states "±1" with positive and negative axial dipolar kernels have a relatively close energy difference $G_0 - G_{\pm 1} \approx + 1.2 \times 10^{-20}$ J corresponding to 2.9 $k_B T$ at 298 K. The manifold degeneracy and energy proximity open possible applications of core-shell nanoparticles and their ensembles as multi-bit memory and related multi-value logic units [84].

To imagine a multi-bit memory cell, let us consider an ensemble of non-interacting core-shell nanoparticles with a vortex polarization placed in a *soft matter environment* (e.g. wax, liquid crystal, or colloid) with a viscosity that is strongly temperature-dependent around the working (i.e. room) temperature. The nanoparticles can freely rotate and move in the liquid soft matter, and are prevented from rotating when the matter becomes solid. The ferroelectric transition temperature of the particle core (~ 390 K) is much higher than the soft matter melting temperature. The direction of the crystallographic axes in a given ferroelectric core is random with respect to the electrodes, which coat opposite sides of the cell, and the core-shell nanoparticles sit between these electrodes. Since the equilibrium vortex axis should coincide with one of the crystallographic directions [100], [010], or [001], a given nanoparticle in the composite can be in any of three types of vortex states, with an axial kernel (two states "±1") and without one (state "0"). The resulting polarization of a nanocomposite can be presented as a sum of these states,

$$|P\rangle = \sum_{ijk} C_{ijk} |ijk\rangle. \tag{7}$$

The real number $C_{ijk}$ is a relative fraction of the state $|ijk\rangle$ in the composite, $\sum_{ijk} C_{ijk}^2 = 1$ and $0 \leq C_{ijk} \leq 1$. The first subscript $i$=1, 2, 3 designates the vortex axis, the second subscript $j$=−1, 0, +1 indicates the kernel "**color**" (i.e. polarization sign "1" or "−1" in the kernel, "0" without a kernel), and the third subscript k= "*l*" or "*r*" corresponds to the "chirality", defined as a clockwise or counterclockwise direction of the vortex polarization rotation. For a given nanoparticle the coefficients can be interpreted as probabilities. The kernel characteristic "color" introduced here is similar to the spin-flavor characteristics used to describe specific states of elementary particles in magnetic fields.

Note that cannot distinguish between some of the 18 states that are calculated using Eq.7 and those measured using current atomic force microscopy (C-AFM) to determine the of local conductivity. Actually, imagine that a single-layer of spherical core-shell nanoparticles with the vortex polarization of the ferroelectric core is placed in a plane capacitor with sizes that are slightly larger than the particle diameter $2(R + \Delta R)$ (see **Fig. 12**). Since the BaTiO$_3$ core is regarded as a dielectric, and the thin SrTiO$_3$ shell and the soft matter surrounding the core-shell particle are low conducting band-gap materials, the optimal conductance path depends on how close the electrodes are to the conductive vortex axis [see dotted light blue path in **Fig.12(b)**]. The resistance $\rho_L$ of the serial connection "core-shell nanoparticle + soft matter" can be estimated from the expression $\rho_L(\alpha) = 2R[(1 - \cos\alpha)\rho_W + \rho_k]$, where $\alpha$ is the angle between the kernel axis and the normal to the electrodes, $\rho_W$ is the specific resistivity of the soft matter, and $\rho_k$ is the specific resistivity of kernel domain walls. To observe the C-AFM contrast between the soft matter and the nanoparticle, the strong inequality $\rho_W \gg \rho_k$ should be valid. Hence, the measurements of local conductivity can distinguish only the projection of the kernel axis on the electrode plane, proportional to $\cos(\alpha)$. The states with $\alpha = \pm\frac{\pi}{2}$ and a simple vortex without a kernel are indistinguishable [see **Fig.12(a)**], as well as the states with different signs of the kernel polarization [see **Fig.12(b)-(c)**]. The conclusion is true for non-interacting nanoparticles only, i.e. when a dipole-dipole interaction between the particles is negligibly small. The nanoparticles in a dense ensemble tend to align the dipole moments of their kernels locally antiparallel; therefore, $\langle P \rangle = 0$ for the macroscopic state resulting in a minimal dipole-dipole energy [such alignment of the states "+1" and "-1" is shown in **Fig.12(c)**].

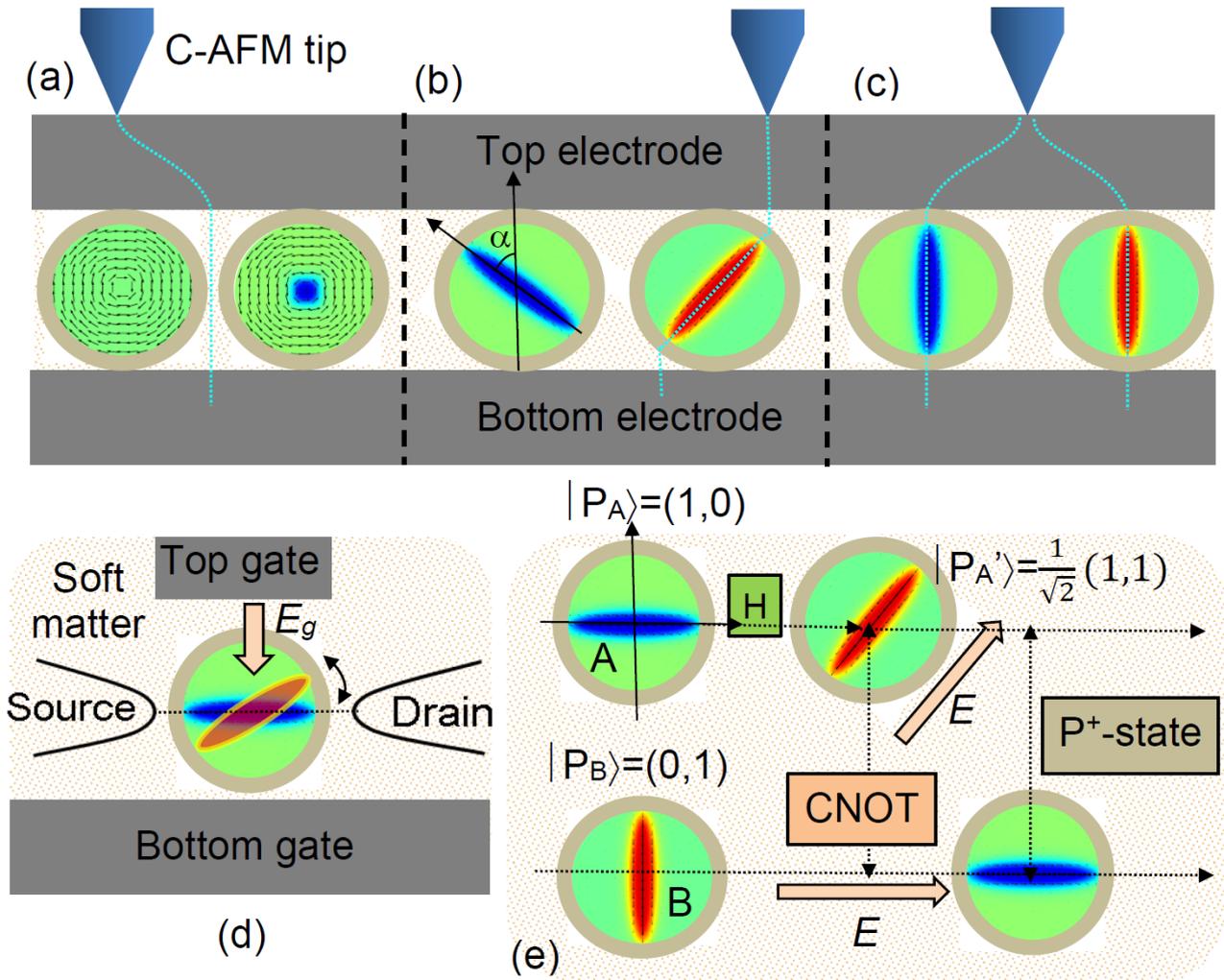

**FIGURE 12.** Spherical core-shell nanoparticles with a vortex polarization and axial dipolar kernel are placed in a plane capacitor with a thickness slightly greater than the sphere diameter. **(a-c)** If the kernel is conductive

the system can be in a low conductivity state (a), intermediate state (b), or high conductivity state (c). **(d)** Nano-FET with a mobile rotating channel. **(e)** The polarization states, $|P_A\rangle$ and $|P_B\rangle$, of two core-shell nanoparticles, "A" and "B". These states coupling or "entanglement" after the application of "H" and "CNOT" operations.

A single core-shell nanoparticle, whose polarization vortex has a dipolar kernel, placed in a soft matter environment, can be considered as a candidate for a nanosized field effect transistor (**nano-FET**) containing a "rotating" channel with an angular dependence of the local resistivity $\rho_L(\alpha)$. Actually, the voltage applied between the FET gates can rotate or shift the particle (in order to rotate/move the kernel axis) [see **Fig.12(d)**], and so the gate field $E_g$ controls the channel conductivity $\sigma_L(\alpha) = \frac{1}{2R[(1-cos\alpha)\rho_W + \rho_k]}$. Note, that the soft matter softening or hardening with temperature can provide an additional temperature control of the channel conductivity. The vortex stability and kernel rotation feasibility are advantages of the nano-FET operation. Its drawback is a relatively low operation speed due to the sluggishness of sphere rotation and/or a translational motion towards the gate.

Another interesting aspect is that a "classical" behavior of the vortex axis (with or without a conductive kernel) can simulate a "qubit" at room temperature, since it formally appeared that some basic properties of qubits necessary for a quantum computation [85, 86] can be simulated by the vortex states "±1" revealed in this work. These possibilities are discussed below, with a strict precaution that we consider an *imitation* of quantum behavior by classical core-shell nanoparticles decoupled or coupled by non-local interactions.

### A. Continuity of polarization and conductivity states over a spherical surface

For a core-shell nanoparticle placed in a liquid medium an almost free rotation of a vortex axis is possible. The rotation is characterized by two spherical angles, θ and ψ. Since the absolute value of the kernel polarization is conserved at zero external field (**E**=0), the polarization vector can be expanded in orthogonal basis vectors, $|P\rangle = A|P_A\rangle + B|P_B\rangle$. Ket vectors $|P_A\rangle = (1,0)$ and $|P_B\rangle = (0,1)$ are the orthogonal basis of one-particle polarization states [see **Fig. 12(e)**]. The representation formally coincides with the Bloch sphere representation of a qubit [85, 86]. The complex coefficients $A = cos\left(\frac{\theta}{2}\right)$ and $B = sin\left(\frac{\theta}{2}\right)e^{i\psi}$ are probability amplitudes for the superposition state, thus $|A|^2 + |B|^2 = 1$; and the local conductivity is sensitive to the superposition. Since the vortex state "0" without a kernel ($A = B = 0$) falls into the center of the sphere, the situation for all three polarization states "0" and "±1" is more complex than the Bloch representation.

The general state of a qubit can be a coherent superposition of two binary "pure" states "0" or "1", while the classical bit can be only either in the state "0" or in the state "1". Whereas a measurement of a classical bit would not disturb its state, a measurement of a qubit would irreversibly destroy its coherence and superposition state (so-called "collapse" of the wave function [85, 86]). Actually, we cannot measure the direction of the vortex axis in a given nanoparticle by application of a voltage between the electrodes, because the particle would rotate in the electric field, and after the rotation time (that is on the order of the Landau-Khalatnikov relaxation time $\tau_K \sim (10^{-9} - 10^{-6})$ s, see **Fig. 9**) the axis of a dipolar kernel aligns parallel to the **E**-field, i.e. perpendicular to the electrodes. The higher the voltage, the faster the vortices are destroyed.

The "collapse" of the core states can be avoided by freezing the soft matter. In this case, the nanoparticle axes cannot rotate and we can measure a "snapshot" of the vortex axes directions (at least hypothetically) in a solid nanocomposite.

### B. Long-range interaction between core-shell nanoparticles

Multiple qubits can exhibit quantum entanglement unlike classical bits. Quantum entanglement is a nonlocal property of two or more qubits that allows a set of qubits to express a higher correlation than is possible in classical systems [85, 86]. Due to the partial screening of the core spontaneous polarization by the paraelectric shell and dielectric (or wide-gap semiconducting) surrounding medium, the core-shell nanoparticles interact via long-range electrostatic forces between their dipolar kernels. The interaction is essentially nonlocal due to the integral nature of depolarization factors and strong spatial dispersion of the effective dielectric permittivity. Simply speaking, the system consisting of two (or more) nanoparticles tends to align the dipole moment of their kernels in an antiparallel alignment to minimize the electrostatic energy. The electrostatic interaction between the particles becomes very weak for an effective (i.e. almost complete) screening of their ferroelectric polarization, and its strength can be controlled by the distance between the particles [43].

Two core-shell nanoparticles, "A" and "B", shown in **Fig. 12(e)**, can be described by the direct product of $|P_A\rangle$ and $|P_B\rangle$ states, with orthogonal basis $|AA\rangle = (1000)$, $|AB\rangle = (0100)$, $|BA\rangle = (0010)$, and $|BB\rangle = (0001)$ in a 4-dimensional Hilbert space. To imitate maximally entangled Bell state P$^+$, one can start from detangled $|P_A\rangle$ and $|P_B\rangle$ states. At first one can apply a Hadamard transformation $H = \frac{1}{\sqrt{2}}\begin{pmatrix} 1 & 1 \\ 1 & -1 \end{pmatrix}$ to $|P_A\rangle$ and obtain $|P_{A'}\rangle = \frac{1}{\sqrt{2}}(1,1)$ as the result. Next, the application of a controllable logic operation "NOT" (i.e. "**CNOT**") is realized by a unitary matrix $U_{CNOT} = \begin{pmatrix} 1000 \\ 0100 \\ 0001 \\ 0010 \end{pmatrix}$ [87]. Applying $U_{CNOT}$ to the direct product of $|P_{A'}\rangle$ and $|P_B\rangle$ states, $\frac{1}{\sqrt{2}}(|AA\rangle + |BA\rangle)$, gives the Bell state $|P_+\rangle = \frac{1}{\sqrt{2}}(|AA\rangle + |BB\rangle)$ as a result. To realize the action of $H$ on the $|P_A\rangle$ state we can apply a sinusoidal pulse of an electric field directed along the axis [101] of the core-shell nanoparticle "A". The field rotates $|P_A\rangle$ vector 45 degrees, and also changes the kernel color (i.e. the sign of

polarization) after the sinusoidal pulse (see e.g. **Fig. 7**). Only after application of $H$ to nanoparticle "A", one can apply another electric field pulse to nanoparticle "B", where the aim is to rotate its $|P_B\rangle$ vector 90 degrees and change its kernel color.

Since it seems possible to imitate a quantum CNOT operation using the considered multi-degenerate ferroelectric vortex states, it may be reasonable to further elaborate on the states possibilities for the imitation of quantum computing [85, 86]. However, one should realize that the long-range electrostatic (or magnetostatic) interaction between the core-shell ferroelectric (or ferromagnetic) nanoparticles has no similarities with "true" entanglement of e.g. photons, because the photons can be entangled at macroscopic distances [86], whereas the nanoparticles cannot be entangled at such distances due to the attenuation of electrostatic or magnetostatic fields. Also, one should realize that the rotation time of ferroelectric or ferromagnetic particle is small, but much longer than the time of almost instant quantum interactions.

## VI. SUMMARY

In the framework of Landau-Ginzburg-Devonshire approach coupled with electrostatic equations, we studied the influence of irrotational external electric fields on the formation, evolution, and relaxation of polarization vortices in spherical ferroelectric nanoparticles. In particular, we performed FEM of the zero-field and in-field polarization evolution in a ferroelectric $BaTiO_3$ core covered with a "tunable" paraelectric $SrTiO_3$ shell placed in a dielectric polymer or liquid medium. The role of the tunable paraelectric shell is to provide an effective screening of the core polarization.

A stable two-dimensional vortex was formed in the core after a zero-field relaxation of a random or poly-domain distribution of polarization, with the vortex axis directed along one of the core crystallographic axes. In order to analyze field-induced changes of the vortex structure, a sinusoidal pulse of a homogeneous electric field was applied, and its period, strength, and direction were varied. A small quasi-static field induced an axial kernel inside the vortex, which formed of a prolate nanodomain with near-homogeneous spontaneous polarization.

Similar to ferromagnetic vortexes, the kernel of a ferroelectric vortex acts like a dipole and opens the possibility to rotate the vortex axis by a small quasi-static homogeneous electric field. Due to the strong ferroelectric anisotropy, which effectively prevents the rotation of the dipolar kernel and tends to align it parallel to the one of crystallographic directions, the in-field orientation of the kernel is conditioned by the minimum of dipolar and anisotropy energies. As the kernel is coupled to the azimuthal vortex, and the vortex plane is perpendicular to the kernel axis, it is rotated with the kernel. The kernel increases in size with an increasing electric field, and eventually the nanoparticle becomes single-domain.

Quite unexpectedly, we revealed the appearance of Bloch point structures during the switching process. In the case of ferromagnetic materials, Bloch points are typically formed at the center of special types of three-dimensional vortices, where the transition between oppositely oriented regions cannot proceed continuously. Here we found analogous situations leading to a Bloch point with a vanishing polarization ($P = 0$). In the considered case, the in-field evolution of the polarization distribution contains two diametrically opposite Bloch points located at the core surface.

After the electric field is removed, the polarization vortex recovers spontaneously, but its structure, axis orientation, and polarization rotation direction can be different from the initial state. As a rule, the final state is a three-dimensional polarization vortex with an axial dipolar kernel, whose energy is less than the energy of the initial two-dimensional "empty" vortex by ~ 3 $k_BT$ at room temperature. The nature of this counterintuitive result is a significant reduction of the gradient energy in the axial region of the vortex by the formation of a kernel, which is only partly compensated by an increase in the energy of the depolarization field for a vortex with a prolate single-domain kernel. The relatively small difference in energies means the possible coexistence of both vortex states.

We also calculated the torque and the electrostatic forces acting on the core-shell nanoparticle placed in an applied electric field. The torque acting on the polarization vortex with an axial kernel tends to rotate the nanoparticle in such way that the vortex axis becomes parallel to the field direction. The vortex (with or without a kernel) is electrostatically neutral, and therefore the nanoparticle does not move in a homogeneous field. Because of the kernel dipole moment, a particle motion can be observed in electric fields with a strong spatial gradient.

The vortex states possess a sixfold degeneracy regarding the vortex orientation and circulation, consisting of three crystallographic axis directions multiplied by the clockwise and counterclockwise directions of the polarization rotation along the axis. As two different "colors" of vortex states can exist, with and without an axial dipolar kernel (the latter with a two-fold degeneracy in terms of the polarization direction on the kernel), the number of states can further be multiplied by three. This multitude of vortex states in a single core makes core-shell nanoparticles and their ensembles promising candidates for multi-bit memory and related logic units. The opportunity to use the rotation of a vortex kernel, which is possible in the case of core-shell nanoparticles placed in a soft matter matrix with controllable viscosity, may be used to imitate qubit properties.

**APPENDIX A** contains a mathematical formulation of the problem in the framework of Landau-Ginzburg-Devonshire theory, and parameters of BaTiO$_3$ (core) and SrTiO$_3$ (shell) materials used in FEM.

**APPENDIX B** contains the approximate solutions of Euler-Lagrange equations and fitting parameters for vortex states.

**APPENDIX C** contains the details of torque and electrostatic force calculations.

**Acknowledgements.** A.N.M. acknowledges EOARD project 9IOE063 and related STCU partner project P751. V.Y.R. acknowledges the support of COST Action CA17139.

**Authors' contribution.** A.N.M. generated the research idea, stated the problem, and performed analytical calculations jointly with V.Yu.R. and V.T. E.A.E. and Y.M.F. wrote the codes and performed numerical calculations. R.H. analyzed of the polarization vortex structures and generated the kernel concept. A.N.M, R.H., D.R.E. and V.Y.R. proposed possible applications. A.N.M. and R.H. wrote the manuscript draft. R.H., V.Yu.R., and D.R.E. worked intensively on the results interpretation and manuscript improvement.

# Supplementary Material to

# "Electric Field Control of Three-Dimensional Vortex States in Core-Shell Ferroelectric Nanoparticles"

### APPENDIX A. Mathematical Formulation of the Problem

We consider a ferroelectric nanoparticle core of radius $R$ with a three-component ferroelectric polarization $\boldsymbol{P}$ directed along one of the crystallographic axes. The core is regarded as insulating, without any free charges. It is covered with a semiconducting tunable shell of thickness $\Delta R$ that is characterized by the relative dielectric permittivity tensor $\varepsilon_{ij}^{S}$. The core-shell nanoparticle is placed in a dielectric medium (polymer, gas, liquid, air, or vacuum) with an effective dielectric permittivity, $\varepsilon_e$. The word "effective" implies the presence of other particles in the medium, which can be described in an effective medium approach. For the sake of clarity, we consider the medium as being isotropic and temperature-independent, i.e. $\varepsilon_{ij}^{e} = \delta_{ij}\varepsilon_e$, in contrast to anisotropic and/or tunable shells. The considered physical model corresponds to a nanocomposite consisting of core-shell nanoparticles in a dielectric medium, with a small volume fraction of ferroelectric nanoparticles (less than 10%) in the composite. The core-shell geometry is shown in **Fig. 1** of the main text.

Below we assume that the shell is soft enough not to affect the strain and stress in the ferroelectric core. The role of the shell is to modify and affect the electrostatics only, and so the elastic part of the problem is the same as in our previous study [1] of a ferroelectric core alone.

Since the ferroelectric polarization contains background and soft mode contributions, the electric displacement vector has the form $D = \varepsilon_0\varepsilon_b E + P$ inside the core. In this expression $\varepsilon_b$ is a relative permittivity of the core background unrelated with the soft mode [2], and $\varepsilon_0$ is the universal dielectric constant, and $\boldsymbol{P}$ is a ferroelectric polarization containing the spontaneous and field-induced contributions, $\boldsymbol{P} = \boldsymbol{P_S} + \varepsilon_0\hat{\chi}_f\boldsymbol{E} + \varepsilon_0\hat{\chi}_{ff}\boldsymbol{E^3} + \varepsilon_0\hat{\chi}_{fff}\boldsymbol{E^5} + ..$, where $\boldsymbol{P_S}$ is the spontaneous polarization at $\boldsymbol{E}$=0. Note that the expression $D = \varepsilon_0\varepsilon_b E + P$ is different from the usual textbook definition, $\boldsymbol{D} = \varepsilon_0\boldsymbol{E} + \boldsymbol{P}$, where $\boldsymbol{P}$ is the total polarization. Usually $4 < \varepsilon_b < 10$, and so $\varepsilon_b$ can be significantly smaller than the linear susceptibility $\chi_f$ related to the soft ferroelectric mode, since as a rule $\chi_f > 30$. In the case of linear response to a small external electric field the displacement is $\boldsymbol{D} \approx \varepsilon_0\hat{\varepsilon}_f\boldsymbol{E} + \boldsymbol{P_S}$, where $\hat{\varepsilon}_f = \hat{\chi}_f + \hat{\varepsilon}_b$. $D_i = \varepsilon_0\varepsilon_{ij}^{S}E_j$ in the shell and $D_i = \varepsilon_0\varepsilon_e E_i$ in the isotropic effective medium.

The electric field components $E_i$ are derived from the electric potential φ in a conventional way, $E_i = -\partial\varphi/\partial x_i$. The potential φ satisfies the Poisson equation in the ferroelectric core (subscript "$f$"):

$$\varepsilon_0\varepsilon_b\left(\frac{\partial^2}{\partial x_1^2} + \frac{\partial^2}{\partial x_2^2} + \frac{\partial^2}{\partial x_3^2}\right)\varphi_f = \frac{\partial P_i}{\partial x_i}, \qquad 0 \le r \le R, \qquad (A.1a)$$

and Debye-type equation in the shell (subscript "$s$"):

$$\varepsilon_0 \frac{\partial}{\partial x_i} \left( \varepsilon_{ij}^S \frac{\partial \varphi_s}{\partial x_j} \right) = -\varepsilon_0 \frac{\varphi_s}{R_d^2}, \qquad R < r < R + \Delta R, \tag{A.1b}$$

where $R_d$ is the "net" screening length of the shell. As mentioned in the introduction, there are several possible ways to change the dielectric permittivity tensor of the shell. In all these cases the external stimuli (electric field, temperature, light, heat, etc.) affect the dielectric properties of the shell, which can influence the spatial distribution of ferro-active ions inside the core via electrostatic interactions and electric boundary conditions.

Outside the shell $\varphi$ satisfies the Laplace equation:

$$\varepsilon_0 \varepsilon_e \left( \frac{\partial^2}{\partial x_1^2} + \frac{\partial^2}{\partial x_2^2} + \frac{\partial^2}{\partial x_3^2} \right) \varphi_e = 0, \qquad r > R + \Delta R, \tag{A.1c}$$

Equations (A.1) are supplemented with the continuity conditions for electric potential and normal components of the electric displacements at the particle surface and core-shell interface:

$$(\varphi_e - \varphi_s)|_{r=R+\Delta R} = 0, \quad \boldsymbol{n}(\boldsymbol{D}_e - \boldsymbol{D}_s)|_{r=R+\Delta R} = 0, \tag{A.1d}$$

$$\left. (\varphi_s - \varphi_f) \right|_{r=R} = 0, \quad \boldsymbol{n}(\boldsymbol{D}_s - \boldsymbol{D}_f)|_{r=R} = 0. \tag{A.1e}$$

Since we do not apply an external field, the potential vanishes either at infinity, $\varphi_e|_{r \to \infty} = 0$, or at the surface of remote electrodes located at the boundaries of the computational region.

The LGD free energy functional $G$ additively includes a Landau expansion on powers of 2-4-6 of the polarization, $G_{Landau}$; a polarization gradient energy contribution, $G_{grad}$; an electrostatic contribution, $G$; as well as elastic, electrostriction, flexoelectric contributions, $G_{es+flexo}$, and surface energy, $G_S$. It has the form [3, 4]:

$$G = G_{Landau} + G_{grad} + G_{el} + G_{es+flexo} + G_S, \tag{A.2a}$$

$$G_{Landau} = \int_{0<r<R} d^3 r \left[ a_i P_i^2 + a_{ij} P_i^2 P_j^2 + a_{ijk} P_i^2 P_j^2 P_k^2 \right], \tag{A.2b}$$

$$G_{grad} = \int_{0<r<R} d^3 r \frac{g_{ijkl}}{2} \frac{\partial P_i}{\partial x_j} \frac{\partial P_k}{\partial x_l}, \tag{A.2c}$$

$$G_{el} = - \int_{0<r<R} d^3 r \left( P_i E_i + \frac{\varepsilon_0 \varepsilon_b}{2} E_i E_i \right) - \frac{\varepsilon_0}{2} \int_{R<r<R+\Delta R} \varepsilon_{ij}^S E_i E_j d^3 r - \frac{\varepsilon_0}{2} \int_{r>R+\Delta R} \varepsilon_{ij}^e E_i E_j d^3 r, \tag{A.2d}$$

$$G_{es+flexo} = - \int_{0<r<R} d^3 r \left( \frac{s_{ijkl}}{2} \sigma_{ij} \sigma_{kl} + Q_{ijkl} \sigma_{ij} P_k P_l + \frac{F_{ijkl}}{2} \left( \sigma_{ij} \frac{\partial P_l}{\partial x_k} - P_l \frac{\partial \sigma_{ij}}{\partial x_k} \right) \right) -$$

$$\int_{R<r<R+\Delta R} d^3 r \frac{s_{ijkl}^S}{2} \sigma_{ij} \sigma_{kl}, \tag{A.1e}$$

$$G_S = \int_{r=R} d^2 r \, a_i^{(S)} P_i^2. \tag{A.1f}$$

The coefficient $a_i$ linearly depends on temperature $T$:

$$a_i(T) = \alpha_T [T - T_C(R)], \tag{A.3a}$$

where $\alpha_T$ is the inverse Curie-Weiss constant and $T_C(R)$ is the ferroelectric Curie temperature renormalized by electrostriction and surface tension. Actually, the surface tension induces additional surface stresses $\sigma_{ij}$ proportional to the surface tension coefficient $\mu$ and equal to $\sigma_{11} = \sigma_{22} = \sigma_{33}|_{r=R} =$ $\frac{-2\mu}{R}$ for a spherical nanoparticle of radius $R$. The stresses affect the Curie temperature and ferroelectric polarization behaviour due to the electrostriction coupling. Thus, the renormalized Curie temperature, $T_C(R)$, acquires the following form [1-3]:

$$T_C(R) = T_C\left(1 - \frac{Q}{\alpha_T T_C}\frac{2\mu}{R}\right) \tag{A.3b}$$

where $T_C$ is a Curie temperature of a bulk ferroelectric. $Q$ is the sum of electrostriction tensor diagonal components that is positive for most ferroelectric perovskites with cubic m3m symmetry in the paraelectric phase, namely $0.004 < Q < 0.04 \mathrm{m^4/C^2}$ [1-4]. Recent experiments tell us that $\mu$ is relatively small, not more than $(2-4)$ N/m for most perovskites.

Tensor components $a_{ij}$ and $a_{ijk}$ are regarded as temperature-independent. The tensor $a_{ij}$ is positively defined if the ferroelectric material undergoes a second order transition to the paraelectric phase and negative otherwise. Higher nonlinear tensor $a_{ijk}$ and gradient coefficients tensor $g_{ijkl}$ are positively defined and regarded as temperature independent. The value $\sigma_{ij}$ is the stress tensor, $s_{ijkl}$ is the elastic compliances tensor, $Q_{ijkl}$ is electrostriction tensor, and $F_{ijkl}$ is the flexoelectric tensor in Eq.(A.2e).

Allowing for the Khalatnikov mechanism of polarization relaxation [5], minimization of the free energy (A.2) with respect to polarization leads to three coupled time-dependent Euler-Lagrange equations for polarization components, $\frac{\delta G}{\delta P_i} = -\Gamma\frac{\partial P_i}{\partial t}$, where the explicit form for a ferroelectric nanoparticle with m3m parent symmetry is:

$$\Gamma\frac{\partial P_1}{\partial t} + 2P_1(a_1 - Q_{12}(\sigma_{22} + \sigma_{33}) - Q_{11}\sigma_{11}) - Q_{44}(\sigma_{12}P_2 + \sigma_{13}P_3) + 4a_{11}P_1^3 + 2a_{12}P_1(P_2^2 + P_3^2) +$$
$$6a_{111}P_1^5 + 2a_{112}P_1(P_2^4 + 2P_1^2P_2^2 + P_3^4 + 2P_1^2P_3^2) + 2a_{123}P_1P_2^2P_3^2 - g_{11}\frac{\partial^2 P_1}{\partial x_1^2} - g_{44}\left(\frac{\partial^2 P_1}{\partial x_2^2} + \frac{\partial^2 P_1}{\partial x_3^2}\right) -$$
$$(g'_{44} + g_{12})\frac{\partial^2 P_2}{\partial x_2\partial x_1} - (g'_{44} + g_{12})\frac{\partial^2 P_3}{\partial x_3\partial x_1} + F_{11}\frac{\partial \sigma_{11}}{\partial x_1} + F_{12}\left(\frac{\partial \sigma_{22}}{\partial x_1} + \frac{\partial \sigma_{33}}{\partial x_1}\right) + F_{44}\left(\frac{\partial \sigma_{12}}{\partial x_2} + \frac{\partial \sigma_{13}}{\partial x_3}\right) = E_1$$

$$\tag{A.4a}$$

$$\Gamma\frac{\partial P_2}{\partial t} + 2P_2(a_1 - Q_{12}(\sigma_{11} + \sigma_{33}) - Q_{11}\sigma_{22}) - Q_{44}(\sigma_{12}P_1 + \sigma_{23}P_3) + 4a_{11}P_2^3 + 2a_{12}P_2(P_1^2 + P_3^2) +$$
$$6a_{111}P_2^5 + 2a_{112}P_2(P_1^4 + 2P_2^2P_1^2 + P_3^4 + 2P_2^2P_3^2) + 2a_{123}P_2P_1^2P_3^2 - g_{11}\frac{\partial^2 P_2}{\partial x_2^2} - g_{44}\left(\frac{\partial^2 P_2}{\partial x_1^2} + \frac{\partial^2 P_2}{\partial x_3^2}\right) -$$
$$(g'_{44} + g_{12})\frac{\partial^2 P_1}{\partial x_2\partial x_1} - (g'_{44} + g_{12})\frac{\partial^2 P_3}{\partial x_3\partial x_2} + F_{11}\frac{\partial \sigma_{22}}{\partial x_2} + F_{12}\left(\frac{\partial \sigma_{11}}{\partial x_2} + \frac{\partial \sigma_{33}}{\partial x_2}\right) + F_{44}\left(\frac{\partial \sigma_{12}}{\partial x_1} + \frac{\partial \sigma_{23}}{\partial x_3}\right) = E_2$$

$$\tag{A.4b}$$

$$\Gamma\frac{\partial P_3}{\partial t} + 2P_3(a_1 - Q_{12}(\sigma_{11} + \sigma_{22}) - Q_{11}\sigma_{33}) - Q_{44}(\sigma_{13}P_1 + \sigma_{23}P_2) + 4a_{11}P_3^3 + 2a_{12}P_3(P_1^2 + P_2^2) +$$
$$6a_{111}P_3^5 + 2a_{112}P_3(P_1^4 + 2P_3^2P_1^2 + P_2^4 + 2P_2^2P_3^2) + 2a_{123}P_3P_1^2P_2^2 - g_{11}\frac{\partial^2 P_3}{\partial x_3^2} - g_{44}\left(\frac{\partial^2 P_3}{\partial x_1^2} + \frac{\partial^2 P_3}{\partial x_2^2}\right) -$$

$$\left(g_{44}' + g_{12}\right)\frac{\partial^2 P_1}{\partial x_3 \partial x_1} - \left(g_{44}' + g_{12}\right)\frac{\partial^2 P_2}{\partial x_3 \partial x_2} + F_{11}\frac{\partial \sigma_{33}}{\partial x_3} + F_{12}\left(\frac{\partial \sigma_{11}}{\partial x_3} + \frac{\partial \sigma_{33}}{\partial x_3}\right) + F_{44}\left(\frac{\partial \sigma_{13}}{\partial x_1} + \frac{\partial \sigma_{23}}{\partial x_2}\right) = E_3$$

$$(A.4c)$$

The Khalatnikov coefficient $\Gamma$ determines the relaxation time of the polarization $\tau_K = \Gamma/|\alpha|$, which typically varies in the range $(10^{-9} – 10^{-6})$ seconds for temperatures far from $T_C$. As argued by Hlinka et al. [6], it is reasonable to assume that $g_{44}' = -g_{12}$.

The boundary condition for polarization at the core-shell interface $r = R$ accounts for the flexoelectric effect:

$$\left(a_i^{(S)}P_i + g_{ijkl}\frac{\partial P_k}{\partial x_l} - F_{klij}\sigma_{kl}\right)n_j\Big|_{r=R} = 0 \qquad (A.5)$$

where **n** is the outer normal to the surface, $i$=1, 2, 3.

Elastic stresses satisfy the equation of mechanical equilibrium in the nanoparticle and its shell,

$$\frac{\partial \sigma_{ij}}{\partial x_j} = 0, \qquad 0 < r < R + \Delta R. \qquad (A.6a)$$

Equations of state follow from the variation of the energy (A.2e) with respect to elastic stress, $\frac{\delta G}{\delta \sigma_{ij}} = -u_{ij}$, namely:

$$s_{ijkl}\sigma_{ij} + Q_{ijkl}P_k P_l + F_{ijkl}\frac{\partial P_l}{\partial x_k} = -u_{ij}, \quad 0 < r < R , \qquad (A.6b)$$

$$s_{ijkl}^S\sigma_{ij} = -u_{ij}, \quad R < r < R + \Delta R , \qquad (A.6b)$$

where $u_{ij}$ is the strain tensor. We will assume that $s_{ijkl}^S \approx s_{ijkl}$ for a "soft" shell. Elastic boundary conditions at the particle core-shell interface $r = R + \Delta R$ are the continuity of the elastic displacement vector and normal stresses.

Below we consider a tunable shell of paraelectric strontium titanate (SrTiO$_3$), which has an isotropic and strongly temperature-dependent dielectric permittivity, $\varepsilon_{ij}^S = \delta_{ij}\varepsilon_s$, with the following expression

$$\varepsilon_s(T) = \frac{1}{\varepsilon_0 \alpha_T T_q^{(E)}}\left(\coth\left(\frac{T_q^{(E)}}{T}\right) - \coth\left(\frac{T_q^{(E)}}{T_0^{(E)}}\right)\right)^{-1} \qquad (A.7)$$

with the Curie-Weiss parameter $\alpha_T$ =0.75×10$^6$ m/(F K) and characteristic temperatures $T_0^{(E)}$ =30 K and $T_q^{(E)}$ =54 K [7]. It should be noted that $\varepsilon_s(T) \approx$3000 at $T$=50 K and $\varepsilon_s(T) \approx$300 at T=298 K allow the spontaneous polarization of the ferroelectric core to be effectively screened by the tunable shell at room and lower temperatures.

**Table AI.** LGD coefficients and other material parameters of a BaTiO$_3$ core covered with a SrTiO$_3$ shell

| Coefficient | Numerical value |
| --- | --- |

| $\varepsilon_{b,s,e}$ | $\varepsilon_b=7$ (core background), $\quad \varepsilon_S=300$ (SrTiO$_3$ shell at 298 K), $\quad \varepsilon_e=10$ (surrounding) |
|---|---|
| $a_i$ (C$^{-2}\cdot$mJ) | $a_1=3.34(T-381)\times10^5$, $\quad\quad$ ($a_1=-2.94\times10^7$ at 298°K) |
| $a_{ij}$ (C$^{-4}\cdot$m$^5$J) | $a_{11}=4.69(T-393)\times10^6-2.02\times10^8$, $a_{12}=3.230\times10^8$, |
| | (at 298°K $a_{11}=-6.71\times10^8$, $a_{12}=3.23\times10^8$) |
| $a_{ijk}$ (C$^{-6}\cdot$m$^9$J) | $a_{111}=-5.52(T-393)\times10^7+2.76\times10^9$, $a_{112}=4.47\times10^9$, $a_{123}=4.91\times10^9$ |
| | (at 298°K $a_{111}=82.8\times10^8$, $a_{112}=44.7\times10^8$, $a_{123}=49.1\times10^8$) |
| $Q_{ij}$ (C$^{-2}\cdot$m$^4$) | $Q_{11}=0.11$, $Q_{12}=-0.043$, $Q_{44}=0.059$ |
| $s_{ij}$ ($\times10^{-12}$ Pa$^{-1}$) | $s_{11}=8.3$, $s_{12}=-2.7$, $s_{44}=9.24$ |
| $g_{ij}$ ($\times10^{-10}$C$^{-2}$m$^3$J) | $g_{11}=1.0$, $g_{12}=(-0.2-+0.3)$, $g_{44}=0.2$ |
| $F_{ij}$ ($\times10^{-11}$C$^{-1}$m$^3$) | $F_{11}=+2.46$, $F_{12}=0.48$, $F_{44}=0.05$ |
| $R_d$ (nm) | >100 nm (shell screening radius) |

## APPENDIX B

Without electrostriction and flexoelectric coupling the explicit expressions of Landau and gradient energy densities have the form:

$$g_{Landau} = a_1(P_1^2 + P_2^2 + P_3^2) + a_{11}(P_1^4 + P_2^4 + P_3^4) + a_{12}(P_1^2 P_2^2 + P_1^2 P_3^2 + P_2^2 P_3^2) +$$
$$a_{111}(P_1^6 + P_2^6 + P_3^6) + a_{112}[P_1^2(P_2^4 + P_3^4) + P_2^2(P_1^4 + P_3^4) + P_3^2(P_2^4 + P_1^4)] + a_{123}P_1^2 P_2^2 P_3^2 \quad \text{(B.1a)}$$

$$g_{grad} = \frac{g_{ijkl}}{2}\frac{\partial P_i}{\partial x_j}\frac{\partial P_k}{\partial x_l}. \quad \text{(B.1b)}$$

The three coupled time-dependent Euler-Lagrange equations (A.4) for the polarization components without electrostriction and flexoelectric coupling are:

$$\Gamma\frac{\partial P_1}{\partial t} + 2a_1 P_1 + 4a_{11}P_1^3 + 2a_{12}P_1(P_2^2 + P_3^2) + 6a_{111}P_1^5 + 2a_{112}P_1(P_2^4 + 2P_1^2 P_2^2 + P_3^4 + 2P_1^2 P_3^2) +$$
$$2a_{123}P_1 P_2^2 P_3^2 - g_{11}\frac{\partial^2 P_1}{\partial x_1^2} - g_{44}\left(\frac{\partial^2 P_1}{\partial x_2^2} + \frac{\partial^2 P_1}{\partial x_3^2}\right) = E_1 \quad \text{(B.2a)}$$

$$\Gamma\frac{\partial P_2}{\partial t} + 2a_1 P_2 + 4a_{11}P_2^3 + 2a_{12}P_2(P_1^2 + P_3^2) + 6a_{111}P_2^5 + 2a_{112}P_2(P_1^4 + 2P_2^2 P_1^2 + P_3^4 + 2P_2^2 P_3^2) +$$
$$2a_{123}P_2 P_1^2 P_3^2 - g_{11}\frac{\partial^2 P_2}{\partial x_2^2} - g_{44}\left(\frac{\partial^2 P_2}{\partial x_1^2} + \frac{\partial^2 P_2}{\partial x_3^2}\right) = E_2 \quad \text{(B.2b)}$$

$$\Gamma\frac{\partial P_3}{\partial t} + 2a_1 P_3 + 4a_{11}P_3^3 + 2a_{12}P_3(P_1^2 + P_2^2) + 6a_{111}P_3^5 + 2a_{112}P_3(P_1^4 + 2P_3^2 P_1^2 + P_2^4 + 2P_2^2 P_3^2) +$$
$$2a_{123}P_3 P_1^2 P_2^2 - g_{11}\frac{\partial^2 P_3}{\partial x_3^2} - g_{44}\left(\frac{\partial^2 P_3}{\partial x_1^2} + \frac{\partial^2 P_3}{\partial x_2^2}\right) = E_3 \quad \text{(B.2c)}$$

For a vortex polarization without a kernel in the form of prolate ellipsoid, where the longer axis of the ellipsoid is directed along [001], the component $P_3 = 0$, and the condition $P = E = 0$ is consistent with $E = 0$ inside the vortex. In a static case the remaining two equations are:

$$2a_1 P_1 + 4a_{11}P_1^3 + 2a_{12}P_1 P_2^2 + 6a_{111}P_1^5 + 2a_{112}P_1(P_2^4 + 2P_1^2 P_2^2) - g_{11}\frac{\partial^2 P_1}{\partial x_1^2} - g_{44}\left(\frac{\partial^2 P_1}{\partial x_2^2} + \frac{\partial^2 P_1}{\partial x_3^2}\right) = 0$$
$$\text{(B.3a)}$$

$$2a_1 P_2 + 4a_{11}P_2^3 + 2a_{12}P_2 P_1^2 + 6a_{111}P_2^5 + 2a_{112}P_2(P_1^4 + 2P_2^2 P_1^2) - g_{11}\frac{\partial^2 P_2}{\partial x_2^2} - g_{44}\left(\frac{\partial^2 P_2}{\partial x_1^2} + \frac{\partial^2 P_2}{\partial x_3^2}\right) = 0$$
$$\text{(B.3b)}$$

The nonlinear coupled equations (B.3) can be solved numerically for specific boundary conditions:

$$\left[a_i^{(S)}P_i + g_{ijkl}\frac{\partial P_k}{\partial x_l}n_j\right]\bigg|_S = 0, \quad \text{(without summation on "}i\text{" in the first term)} \quad \text{(B.4a)}$$

where the explicit form is:

$$\left(a_1^{(S)}P_1 + g_{11}\frac{\partial P_1}{\partial x_1}n_1 + g_{44}\frac{\partial P_1}{\partial x_2}n_2 + g_{44}\frac{\partial P_1}{\partial x_3}n_3 + g_{44}\frac{\partial P_2}{\partial x_1}n_2\right)\bigg|_{r=R} = 0, \quad \text{(B.4b)}$$

$$\left(a_2^{(S)}P_2 + g_{11}\frac{\partial P_2}{\partial x_2}n_2 + g_{44}\frac{\partial P_2}{\partial x_1}n_1 + g_{44}\frac{\partial P_2}{\partial x_3}n_3 + g_{44}\frac{\partial P_1}{\partial x_2}n_1\right)\bigg|_{r=R} = 0. \quad \text{(B.4c)}$$

The integral of motion of the system (B.4) is $g_{Landau} - g_{grad} = const$.

An analytical solution of the nonlinear system (B.3) is possible in several cases for very specific boundary conditions, but not for more general conditions (B.4). Below we consider a hypothetical situation of the stress-free system with two-component polarizations, $P_1(x_3)$ and $P_2(x_3)$, consistent with the absence of a depolarization field.

**Case I. Static one-component and one-dimensional partial solution.** If only one component of the polarization is position dependent, and the other components are zero, the one-dimensional profile of the uncharged domain wall is given by expression

$$P_2(x_3) = \frac{P_S \cdot tanh[(x_3-x_0)/R_c]}{\sqrt{\eta \cdot sech^2[(x_3-x_0)/R_c]+1}}, \qquad P_1(x_3) = 0, \quad \text{(B.5a)}$$

where $P_S$ is the spontaneous polarization, $x_3 - x_0$ is the distance from center of the domain wall plane, and $2 R_c$ is the domain wall width [8]. For second-order ferroelectrics $P_S^2 = -a_1/(2a_{11})$ and $\eta = 0$, while for first-order ferroelectrics $P_S^2 = \left(\sqrt{a_{11}^2 - 4a_1a_{111}} - a_{11}\right)/2a_{111}$ and the dimensionless parameter $\eta = 2(a_1 + a_{11}P_S^2)/(4a_1 + a_{11}P_S^2)$ is positive. The correlation length is $R_c = \sqrt{g_{44}/(a_1 + 3a_{11}P_S^2 + 5a_{111}P_S^4)}$. The expression (B.5a) describes a 180-degree Ising-type uncharged domain wall (see **Fig. B1**). Since $sech^2[x] = 1 - tanh^2[x]$, it can be more convenient to expand Eq.(B.5a) in series of $tanh[(x_3 - x_0)/R_c]$ near the wall plane.

The solution (B.5) should fulfill the boundary condition (B.4a), which acquires the form:

$$\left(a_1^{(S)}P_1 + g_{44}\frac{\partial P_1}{\partial x_3}n_3\right)\bigg|_{r=R} = 0. \quad \text{(B.5b)}$$

As one can see the boundary condition (B.5b) is incompatible with the solution (B.5a) at the spherical surface $r=R$. However, a rather small incompatibility corresponds to the specific case $a_1^{(S)} = 0$, $|x_0| \ll R$ and $R \gg R_c$, since $sech^2[(R - x_0)/R_c] \ll 1$ for this case. Allowing for $R_c$ to be of the order of a lattice constant, considering that the core radius should be not less than $(10 - 20)$ lattice constants for the applicability of continuous LGD approach, the solution (B.5a) can be regarded as a relevant trial function for the case $a_1^{(S)} = 0$.

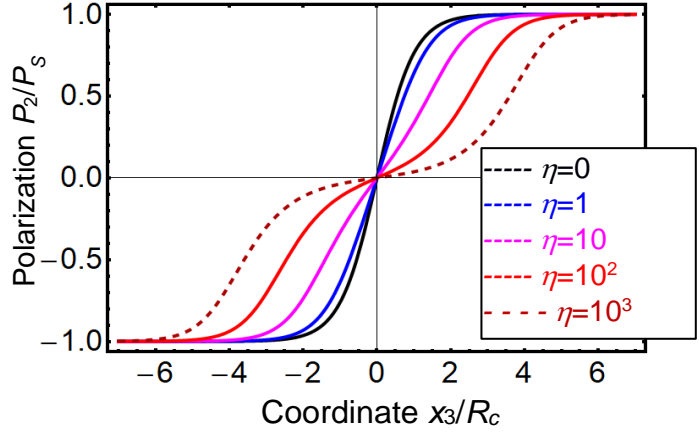

**FIGURE B1.** Profile of the polarization component $P_2/P_S$ across the domain wall, given by Eq.(B.5) for different values of $\eta=0$ (black curve), 1 (blue curve), 10 (magenta curve), $10^2$ (red curve), and $10^3$ (dashed curve).

**Case II. Static two-component and one-dimensional partial solution.** For ferroelectrics with a second order phase transition ($a_{11} > 0$ and $a_{111} = a_{112} = 0$), the one-dimensional profile of uncharged domain walls satisfies Eqs.(B.3), which can be simplified as follows:

$$2a_1 P_1 + 4a_{11} P_1^3 + 2a_{12} P_1 P_2^2 - g_{44} \frac{\partial^2 P_1}{\partial x_3^2} = 0, \tag{B.6a}$$

$$2a_1 P_2 + 4a_{11} P_2^3 + 2a_{12} P_2 P_1^2 - g_{44} \frac{\partial^2 P_2}{\partial x_3^2} = 0. \tag{B.6b}$$

For the specific case $a_{12} = 6a_{11}$, the partial solution of Eqs.(B.6) is [9]:

$$P_1(x_3) = \frac{P_S}{2}\left(tanh\left[\frac{x_3-x_a}{\sqrt{2}R_c}\right] + tanh\left[\frac{x_3-x_b}{\sqrt{2}R_c}\right]\right) \equiv \frac{P_S \cdot sinh[R_0/R_c]}{cosh[R_0/R_c]+cosh[(x_3-x_0)/R_c]}, \tag{B.7a}$$

$$P_2(x_3) = \frac{P_S}{2}\left(tanh\left[\frac{x_3-x_a}{\sqrt{2}R_c}\right] - tanh\left[\frac{x_3-x_b}{\sqrt{2}R_c}\right]\right) \equiv \frac{P_S \cdot sinh[(x_3-x_0)/R_c]}{cosh[R_0/R_c]+cosh[(x_3-x_0)/R_c]}, \tag{B.7b}$$

where $P_S = \sqrt{-a_1/(2a_{11})}$ is the spontaneous polarization ($a_1 < 0$, $x_3 - x_0$ is the distance from the center of the domain wall plane, the correlation length is $R_c = \sqrt{-g_{44}/(2a_1)}$, and $R_0 \sim (x_a - x_b)$ is an arbitrary constant. For the particular case $R_0 = 0$, we obtain $P_1(x_3) = 0$ and $P_2(x_3) = P_S tanh[(x_3-x_0)/R_c]$. For a nonzero $R_0$ the profile (B.7) is an uncharged Ising-Bloch type domain wall (two "rotational" 180-degree c-domains separated by an a-domain, see **Fig. B2**). The wall energy is $R_0$-independent [9].

Note that the boundary condition (B.5) imposed at the spherical surface $r=R$ weakly affects the solution (B.7) for the specific case $a_1^{(S)} = 0$, $|x_0| \ll R$ and $R \gg R_c$. Assuming that these inequalities are valid, the solution (B.7) can be considered as a relevant trial function.

For the three-component solution the sum of three (or more) tanh-functions can be used as a trial function.

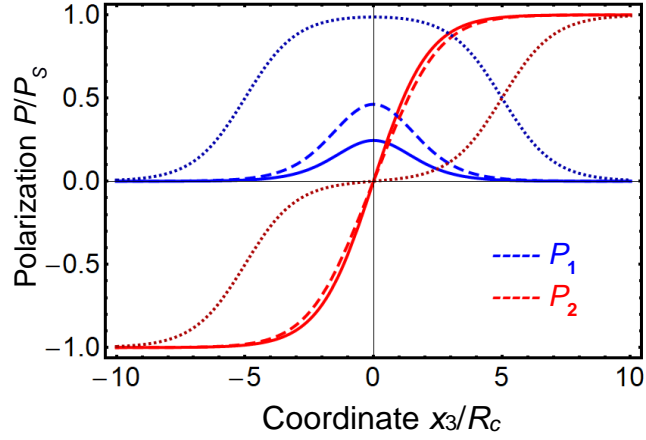

**FIGURE B2.** Profile of polarization components, $P_1/P_S$ (blue curves) and $P_2/P_S$ (red curves), across the domain wall, given by Eq. (B.7) for different values of $R_0/R_c$=0.5 (solid curves), 1 (dashed curves) and 5 (dotted curves).

The static solution of the nonlinear coupled Eqs. (B.3) can be found by a variational principle. Since both partial solutions (B.5) and (B.7) are expressed via hyperbolic functions, and because Figures B1 and B2 look very similar for $P_2$ profiles, we use hyperbolic functions as the basis for the serial expansion of the trial functions:

$$P_k(r,\psi,\theta) = \sum_{i=1}^{3} a_i(\psi,\theta) tanh\left(\frac{r - b_i(\psi,\theta)}{R_i(\psi,\theta)}\right), \qquad k = r, \theta, \psi . \qquad (B.8)$$

The trial functions should be substituted into the energy (B.1b-c), integrated over the particle volume, and minimized in order to determine the unknown coefficients.

Since the equality $\frac{1}{r^2}\frac{\partial}{\partial r}\left(r^2\frac{\partial P_r}{\partial r}\right) = \frac{-1}{r sin\theta}\left(\frac{\partial P_\psi}{\partial \psi} + \frac{\partial(P_\theta sin\theta)}{\partial \theta}\right)$ should be valid for $P(r,\psi) = 0$, we obtain that the first-order harmonic correction to $P_\psi(r)$ is proportional to $sin(\theta)cos(4\psi)$.

**Table BI.** Fitting parameters for the polarization components of the initial vortex at $\theta = \pi/2$

| $P_\alpha$ | $\psi$ | $a_1$ (C/m²) | $b_1$ (nm) | $R_1$ (nm) | $a_2$ (C/m²) | $b_2$ (nm) | $R_2$ (nm) | $a_3$ (C/m²) | $b_3$ (nm) | $R_3$ (nm) |
|---|---|---|---|---|---|---|---|---|---|---|
| $P_r$ | $0, \pi/4$ | 0 | N/A | N/A | 0 | N/A | N/A | 0 | N/A | N/A |
| $P_r$ | $\pi/8, 3\pi/8$ | 0.033 | 1.10 | 1.90 | 0.011 | 0.90 | 0.51 | 2.322 | 3.94 | 342.6 |
| $P_\psi$ | $0$ | 0.296 | 0 | 0.75 | $-0.035$ | $-1.93$ | 6,69 | 0 | N/A | N/A |
| $P_\psi$ | $\pi/8, 3\pi/8$ | 0.275 | 0 | 0.73 | 0.009 | 8.71 | 2.86 | 0 | N/A | N/A |
| $P_\psi$ | $\pi/4$ | 0.290 | 0 | 0.85 | 0.028 | 3.89 | 2.48 | $-0.029$ | 1.19 | 0.84 |
| $P_\theta$ | N/A | 0 | N/A | N/A | 0 | N/A | N/A | 0 | N/A | N/A |

*N/A - non-applicable

**Table BII**. Fitting parameters for the polarization components of the final vortex at $\theta = \pi/2$

| $P_\alpha$ | $\psi$ | $a_1$ (C/m$^2$) | $b_1$ (nm) | $R_1$ (nm) | $a_2$ (C/m$^2$) | $b_2$ (nm) | $R_2$ (nm) | $a_3$ (C/m$^2$) | $b_3$ (nm) | $R_3$ (nm) |
|---|---|---|---|---|---|---|---|---|---|---|
| $P_r$ | $0, \pi/4$ | 0 | N/A | N/A | 0 | N/A | N/A | 0 | N/A | N/A |
| $P_r$ | $\pi/8, 3\pi/8$ | –0.125 | 5.12 | 16.83 | 0.039 | 1.25 | 0.69 | 0 | N/A | N/A |
| $P_\psi$ | 0 | 0.183 | 0.96 | 0.81 | 0.274 | –0.83 | 0.71 | –0.192 | –0.82 | 2.00 |
| $P_\psi$ | $\pi/8, 3\pi/8$ | 0.257 | –7.36 | 7.10 | 0.238 | 0.79 | 1.09 | –0.216 | –0.58 | 2.33 |
| $P_\psi$ | $\pi/4$ | 0.125 | 0.91 | 1.07 | 0.174 | –3.69 | 6.73 | 0 | N/A | N/A |
| $P_\theta$ | 0 | 0.150 | 1.65 | 0.93 | –0.140 | –2.39 | 1.72 | 0 | N/A | N/A |
| $P_\theta$ | $\pi/8, 3\pi/8$ | 0.150 | 1.75 | 0.98 | –0.139 | –1.48 | 1.73 | 0 | N/A | N/A |
| $P_\theta$ | $\pi/4$ | 0.146 | 1.88 | 1.02 | –0.138 | –2.75 | 1.83 | 0 | N/A | N/A |

*N/A - non-applicable

# APPENDIX C

Note that it is easy to check that $\overline{P} = \int_V \boldsymbol{P} dV = 0$, if $div\boldsymbol{P} = 0$ and $P_n = 0$ at the particle surface. Actually $\overline{P}\boldsymbol{c} = \int_V (\boldsymbol{cP}) dV$, where $\boldsymbol{c} \equiv grad(\boldsymbol{rc})$ is an arbitrary constant vector. Then $\boldsymbol{P}\,grad(\boldsymbol{rc}) = div\big((\boldsymbol{rc})\boldsymbol{P}\big) - (\boldsymbol{rc})\,div\boldsymbol{P} = div\big((\boldsymbol{rc})\boldsymbol{P}\big)$ and so $\overline{P}\boldsymbol{c} = \int_S (\boldsymbol{rc}) P_n dS = 0$. Since vector $\boldsymbol{c}$ is arbitrary, the equality $\overline{P}\boldsymbol{c} = 0$ leads to $\overline{P} = 0$.

## C.1. Torque Calculation

The torque **M** acting on a particle in a constant homogeneous field $\boldsymbol{E}^{ext}$ is equal to

$$\boldsymbol{M} = -\big[\boldsymbol{E}^{ext} \times \int_V \boldsymbol{P}(r) d^3 r\big]. \tag{C.1}$$

It is easy to check that $\int_V \boldsymbol{P}(r) d^3 r = 0$ for the "empty" vortex without a kernel, and therefore $\boldsymbol{M} = 0$ for any "true" vortex polarization distribution. Unlike an empty vortex, the vortex with a kernel oriented along z-axis has a non-zero component of the total dipole moment, e.g. $\overline{P_z} = \frac{3\pi}{4R^3}\int_V P_z(\boldsymbol{r}) d^3 r$. Thus $\boldsymbol{M} = \frac{4\pi}{3}R^3\big(E_y^{ext}\boldsymbol{e}_x - E_x^{ext}\boldsymbol{e}_y\big)\overline{P_z}$.

## C.2. Electric Field and Force Calculations

### C.2.1. Homogeneously polarized core-shell nanoparticle in a homogeneous external field

Let us consider a core-shell nanoparticle with a homogeneously polarized core of radius $R$, where the polarization $P_S$ is pointing along the z-axis. The core, shell, and external effective medium are dielectrics. Thus, the electrostatic potential satisfies the Laplace equation in all of the regions:

$$\Delta\varphi_e = 0, \qquad \Delta\varphi_s = 0 \qquad \Delta\varphi_f = 0 , \tag{C.2}$$

where the subscripts "$f$", "$s$", and "$e$" denote the physical quantities related to the ferroelectric core, shell, and external media, respectively. The electric field and displacement vectors are:

$$\mathbf{E}_{f,s,e} = -\nabla\varphi_{f,s,e}, \qquad \mathbf{D}_f = \varepsilon_0\varepsilon_f\mathbf{E}_f + \mathbf{e}_z P_S, \qquad \mathbf{D}_{s,e} = \varepsilon_0\varepsilon_{s,e}\mathbf{E}_{s,e}. \tag{C.3}$$

Here $\boldsymbol{e}_z$ is the unit vector along the axis z. The potential corresponds to a homogeneous electric field $\boldsymbol{e}_z z E^{ext}$ very far away from the particle. The potential and radial displacement are continuous functions at all interfaces:

$$\left.(\varphi_f - \varphi_s)\right|_{r=R} = 0, \quad \left.\mathbf{e}_r \cdot (\mathbf{D}_f - \mathbf{D}_s)\right|_{r=R} = 0, \tag{C.4a}$$

$$\left.(\varphi_s - \varphi_e)\right|_{r=R+\Delta R} = 0, \quad \left.\mathbf{e}_r \cdot (\mathbf{D}_s - \mathbf{D}_e)\right|_{r=R+\Delta R} = 0. \tag{C.4b}$$

Let us switch coordinate systems from Cartesian to spherical coordinates, with the polar axis along the z axis. It is natural to assume that the electrostatic field does not depend on the azimuthal coordinate $\psi$, so the general solution is

$$\varphi_f = -E_f r cos\theta, \quad \varphi_s = -E_s r cos\theta + B_s\frac{cos\theta}{r^2}, \quad \varphi_e = B_e\frac{cos\theta}{r^2} - rcos\theta E^{ext} . \tag{C.5}$$

The radial components of the electric displacement are

$$\mathbf{e}_r\mathbf{D}_f = \varepsilon_0\varepsilon_f E_f\cos\theta + P_S\cos\theta , \quad \mathbf{e}_r\mathbf{D}_s = \varepsilon_0\varepsilon_s\cos\theta\left(E_s + \frac{2B_s}{r^3}\right), \tag{C.6a}$$

$$\mathbf{e}_r \cdot \mathbf{D}_e = \varepsilon_0\varepsilon_e\left(2\frac{cos\theta}{r^3}B_e + cos\theta E^{ext}\right). \tag{C.6b}$$

Substitution of the solution (C.6) into the boundary conditions (C.4) gives a system of linear equations for the unknown coefficients $E_{f,s}$ and $B_{s,e}$. Using their values yields the following expressions:

$$E_f = \frac{9(R+\Delta R)^3\varepsilon_0\varepsilon_e\varepsilon_s E^{ext} - [2R^3(\varepsilon_s-\varepsilon_e)+(R+\Delta R)^3(2\varepsilon_e+\varepsilon_s)]P_S}{\varepsilon_0\left(2R^3(\varepsilon_e-\varepsilon_s)(\varepsilon_s-\varepsilon_f)+(R+\Delta R)^3(2\varepsilon_e+\varepsilon_s)(\varepsilon_f+2\varepsilon_s)\right)}, \tag{C.7a}$$

$$E_s = \frac{3(R+\Delta R)^3\varepsilon_0\varepsilon_e(2\varepsilon_s+\varepsilon_f)E^{ext} - 2R^3(\varepsilon_s-\varepsilon_e)P_S}{\varepsilon_0\left(2R^3(\varepsilon_e-\varepsilon_s)(\varepsilon_s-\varepsilon_f)+(R+\Delta R)^3(2\varepsilon_e+\varepsilon_s)(\varepsilon_f+2\varepsilon_s)\right)}, \tag{C.7b}$$

$$B_s = \frac{R^3(R+\Delta R)^3[(2\varepsilon_e+\varepsilon_s)P_S+3\varepsilon_0(\varepsilon_f-\varepsilon_s)E^{ext}]}{\varepsilon_0\left(2R^3(\varepsilon_e-\varepsilon_s)(\varepsilon_s-\varepsilon_f)+(R+\Delta R)^3(2\varepsilon_e+\varepsilon_s)(\varepsilon_f+2\varepsilon_s)\right)}, \tag{C.7c}$$

$$B_e = \frac{3R^3(R+\Delta R)^3 P_S\varepsilon_s+(R+\Delta R)^3[R^3(\varepsilon_f-\varepsilon_s)(2\varepsilon_s+\varepsilon_e)-(\varepsilon_e-\varepsilon_s)(2\varepsilon_s+\varepsilon_b)(R+\Delta R)^3]\varepsilon_0 E^{ext}}{\varepsilon_0\left(2R^3(\varepsilon_e-\varepsilon_s)(\varepsilon_s-\varepsilon_f)+(R+\Delta R)^3(2\varepsilon_e+\varepsilon_s)(\varepsilon_f+2\varepsilon_s)\right)} . \tag{C.7d}$$

From Eq.(C.7), the electric field inside of the particle core is homogeneous and equal to

$$E_f = \frac{-[2R^3(\varepsilon_s-\varepsilon_e)+(R+\Delta R)^3(2\varepsilon_e+\varepsilon_s)]P_S e_z}{\varepsilon_0\left(2R^3(\varepsilon_e-\varepsilon_s)(\varepsilon_s-\varepsilon_f)+(R+\Delta R)^3(2\varepsilon_e+\varepsilon_s)(\varepsilon_f+2\varepsilon_s)\right)} + \frac{9(R+\Delta R)^3\varepsilon_e\varepsilon_s E^{ext}e_z}{2R^3(\varepsilon_e-\varepsilon_s)(\varepsilon_s-\varepsilon_f)+(R+\Delta R)^3(2\varepsilon_e+\varepsilon_s)(\varepsilon_f+2\varepsilon_s)} \quad \text{(C.8)}$$

Let us set $E^{ext}=0$. Outside the core-shell particle the potential and electric field created by the particle is inhomogeneous and has the form of a point dipole field:

$$\varphi_e = \frac{\mathbf{p}_f \cdot \mathbf{r}}{4\pi\varepsilon_0\varepsilon_e r^3}, \quad \text{(C.9a)}$$

with an effective dipole moment

$$\mathbf{p}_f = \frac{12\pi P_S R^3(R+\Delta R)^3\varepsilon_s\varepsilon_e\mathbf{e}_z}{2R^3(\varepsilon_e-\varepsilon_s)(\varepsilon_s-\varepsilon_f)+(R+\Delta R)^3(2\varepsilon_e+\varepsilon_s)(\varepsilon_f+2\varepsilon_s)}. \quad \text{(C.9b)}$$

If the shell is very thin ($\Delta R \ll R$) or absent ($\Delta R = 0$) the denominator in Eq.(C.9b) is equal to $2(\varepsilon_e-\varepsilon_s)(\varepsilon_s-\varepsilon_f)R^3+(2\varepsilon_e+\varepsilon_s)(\varepsilon_f+2\varepsilon_s)R^3=3\varepsilon_s(2\varepsilon_e+\varepsilon_f)R^3$ and so

$$\mathbf{p}_f \cong \frac{4\pi\varepsilon_e}{2\varepsilon_e+\varepsilon_f}P_S R^3 \mathbf{e}_z. \quad \text{(C.9c)}$$

### C.2.2. Homogeneously polarized core-shell nanoparticle in a point charge field

Next let us consider the case when an inhomogeneous external field is produced by a charged tip (see **Fig. 1c**). For this case the effective point charge model is applicable, and the effective charge is $Q^* \approx C_t U$, where $C_t$ is the capacity, $U$ is the voltage applied to the tip, and $h = \{0,0,h\}$ is the charge location.

For this case the electrostatic potential satisfies the Laplace equation in all of the regions, except for the point charge location:

$$\Delta\varphi_e = -\frac{Q^*}{\varepsilon_0\varepsilon_e}\delta(\mathbf{r}-\mathbf{h}), \qquad \Delta\varphi_s = 0 \qquad \Delta\varphi_f = 0. \quad \text{(C.10)}$$

The electric field, displacement vectors, and boundary conditions are given by Eqs.(C.3) and (C.4), respectively. Using the same spherical coordinates as in subsection C.2.1, the general solution for the electrostatic potential is:

$$\varphi_f = -E_f r\cos\theta + \sum_{n=0}^{\infty} A_n r^n L_n(\cos\theta), \qquad 0 \leq r \leq R, \quad \text{(C.11)}$$

$$\varphi_s = \left(\frac{B_s}{r^2}-E_s r\right)\cos\theta + \sum_{n=0}^{\infty}\left[B_n r^n + C_n\frac{1}{r^{n+1}}\right]L_n(\cos\theta), \qquad R < r < R+\Delta R, \quad \text{(C.12)}$$

$$\varphi_e = B_e\frac{\cos\theta}{r^2} + \sum_{n=0}^{\infty} F_n\frac{1}{r^{n+1}}L_n(\cos\theta) + \frac{Q^*}{4\pi\varepsilon_0\varepsilon_e|\mathbf{r}-\mathbf{h}|}. \qquad r \geq R+\Delta R. \quad \text{(C.13)}$$

where $L_n(t)=\frac{1}{2^n n!}\frac{d^n}{dx^n}[(t^2-1)^n]$ are Legendre polynomials, which are orthogonal, and $\int_0^\pi L_n(\cos\theta)L_m(\cos\theta)\sin\theta d\theta = \frac{2\delta_{nm}}{2n+1}$. The potential is independent on the azimuthal coordinate $\psi$.

The constants $E_{f,s}$ and $B_{s,e}$, which are proportional to $P_S$, are given by Eqs.(C.7) at $E^{ext}=0$:

$$E_f = \frac{-[2R^3(\varepsilon_s-\varepsilon_e)+(R+\Delta R)^3(2\varepsilon_e+\varepsilon_s)]P_S}{\varepsilon_0\left(2R^3(\varepsilon_e-\varepsilon_s)(\varepsilon_s-\varepsilon_f)+(R+\Delta R)^3(2\varepsilon_e+\varepsilon_s)(\varepsilon_f+2\varepsilon_s)\right)}, \quad \text{(C.14a)}$$

$$E_s = \frac{-2R^3(\varepsilon_s - \varepsilon_e)P_S}{\varepsilon_0\big(2R^3(\varepsilon_e - \varepsilon_s)(\varepsilon_s - \varepsilon_f) + (R+\Delta R)^3(2\varepsilon_e + \varepsilon_s)(\varepsilon_f + 2\varepsilon_s)\big)}, \tag{C.14b}$$

$$B_s = \frac{R^3(R+\Delta R)^3(2\varepsilon_e + \varepsilon_s)P_S}{\varepsilon_0\big(2R^3(\varepsilon_e - \varepsilon_s)(\varepsilon_s - \varepsilon_f) + (R+\Delta R)^3(2\varepsilon_e + \varepsilon_s)(\varepsilon_f + 2\varepsilon_s)\big)}, \tag{C.14c}$$

$$B_e = \frac{3R^3(R+\Delta R)^3 P_S \varepsilon_s}{\varepsilon_0\big(2R^3(\varepsilon_e - \varepsilon_s)(\varepsilon_s - \varepsilon_f) + (R+\Delta R)^3(2\varepsilon_e + \varepsilon_s)(\varepsilon_f + 2\varepsilon_s)\big)}. \tag{C.14d}$$

Using that

$$\frac{Q^*}{4\pi\varepsilon_0\varepsilon_e|\mathbf{r}-\mathbf{h}|} = \frac{Q^*}{4\pi\varepsilon_0\varepsilon_e}\begin{cases}\sum_{n=0}^{\infty}\frac{r^n}{h^{n+1}}L_n(\cos\theta) & \text{at } r < h, \\ \sum_{n=0}^{\infty}\frac{h^n}{r^{n+1}}L_n(\cos\theta) & \text{at } r > h,\end{cases} \tag{C.15a}$$

the radial components of the electric displacement are:

$$D_{fr} = \varepsilon_0\varepsilon_f E_f \cos\theta - \varepsilon_0\varepsilon_f \sum_{n=1}^{\infty} A_n n r^{n-1} L_n(\cos\theta), \tag{C.15b}$$

$$D_{sr} = \varepsilon_0\varepsilon_S\left(-2\frac{B_s}{r^3} + E_s\right)\cos\theta - \varepsilon_0\varepsilon_S \sum_{n=0}^{\infty}\left(B_n n r^{n-1} - C_n\frac{n+1}{r^{n+2}}\right)L_n(\cos\theta), \tag{C.15c}$$

$$D_{er} = 2\varepsilon_0\varepsilon_e B_e\frac{\cos\theta}{r^3} - \varepsilon_0\varepsilon_e\sum_{n=0}^{\infty}\left(\frac{Q^*}{4\pi\varepsilon_0\varepsilon_e}\frac{n r^{n-1}}{h^{n+1}} - F_n\frac{n+1}{r^{n+2}}\right)L_n(\cos\theta). \tag{C.15d}$$

After the substitution of expressions (C.11)-(C.15) to the boundary conditions (C.4) and cumbersome algebraic transformations we obtain

$$A_n = \frac{Q^*}{4\pi\varepsilon_0 h^{1+n}}\frac{(1+2n)^2(1+\Delta)^{1+2n}\varepsilon_S}{n(1+n)(\varepsilon_f - \varepsilon_S)(\varepsilon_S - \varepsilon_e) + (1+\Delta)^{1+2n}(\varepsilon_e(1+n) + n\varepsilon_S)(n\varepsilon_f + (1+n)\varepsilon_S)}, \tag{C.16a}$$

$$B_n = \frac{Q^*}{4\pi\varepsilon_0 h^{1+n}}\frac{(1+2n)(1+\Delta)^{1+2n}[n\varepsilon_f + (1+n)\varepsilon_S]}{n(1+n)(\varepsilon_f - \varepsilon_S)(\varepsilon_S - \varepsilon_e) + (1+\Delta)^{1+2n}(\varepsilon_e(1+n) + n\varepsilon_S)(n\varepsilon_f + (1+n)\varepsilon_S)}, \tag{C.16b}$$

$$C_n = \frac{Q^* R^{1+2n}}{4\pi\varepsilon_0 h^{1+n}}\frac{n(1+2n)(\varepsilon_S - \varepsilon_f)(1+\Delta)^{1+2n}}{n(1+n)(\varepsilon_f - \varepsilon_S)(\varepsilon_S - \varepsilon_e) + (1+\Delta)^{1+2n}(\varepsilon_e(1+n) + n\varepsilon_S)(n\varepsilon_f + (1+n)\varepsilon_S)}, \tag{C.16c}$$

$$F_n = \frac{Q^* R^{1+2n}}{4\pi\varepsilon_0 h^{1+n}}\frac{n(1+\Delta)^{1+2n}\big[(1+\Delta)^{1+2n}(\varepsilon_e - \varepsilon_S)(n\varepsilon_f + (1+n)\varepsilon_S) + (\varepsilon_S - \varepsilon_f)(n\varepsilon_e + (1+n)\varepsilon_S)\big]}{n(1+n)(\varepsilon_f - \varepsilon_S)(\varepsilon_S - \varepsilon_e) + (1+\Delta)^{1+2n}(\varepsilon_e(1+n) + n\varepsilon_S)(n\varepsilon_f + (1+n)\varepsilon_S)}, \tag{C.16d}$$

where $\Delta = \frac{\Delta R}{R}$ and $C_0 = F_0 = 0$.

### C.2.3. Interaction of polarized core-shell nanoparticle with a point charge

Let us consider the interaction of the polarized core-shell particle with the point charge $Q^*$, localized at the point with coordinate $r = h$. The corresponding interaction energy is

$$W = \left(Q^*\varphi_p + \frac{1}{2}Q^*\varphi_Q\right)\Big|_{\mathbf{r}=\mathbf{h}} = \frac{Q^*(\mathbf{p}_f\cdot\mathbf{h})}{4\pi\varepsilon_0\varepsilon_e h^3} + \frac{Q^{*2}}{8\pi\varepsilon_0}\sum_{n=1}^{\infty}\frac{R^{1+2n}}{h^{2n+2}}F_n L_n(\cos\theta)\Big|_{\theta=0}. \tag{C.17}$$

Finally, the generalized electrostatic force acting on a point charge is

$$\mathbf{F} = \frac{\partial W}{\partial \mathbf{h}} = \frac{Q^*}{4\pi\varepsilon_0\varepsilon_e h^3}\left[\mathbf{p}_f - 3(\mathbf{p}_f\cdot\mathbf{h})\frac{\mathbf{h}}{h^2}\right] - \frac{Q^{*2}}{4\pi\varepsilon_0}\left(\sum_{n=0}^{\infty}\frac{R^{1+2n}}{h^{3+2n}}G_n\right)\frac{\mathbf{h}}{h} \tag{C.18a}$$

We used the identity $L_n(1) \equiv 1$ in the last term of Eq.(C.18a). The dipole moment,

$$\mathbf{p}_f = \frac{12\pi P_S R^3 (1+\Delta)^3 \varepsilon_s \varepsilon_e \mathbf{e_z}}{2(\varepsilon_e - \varepsilon_s)(\varepsilon_s - \varepsilon_f) + (1+\Delta)^3 (2\varepsilon_e + \varepsilon_s)(\varepsilon_f + 2\varepsilon_s)}. \tag{C.18c}$$

The coefficient $G_n$ is given by the expression

$$G_n = \frac{n(1+n)(1+\Delta)^{1+2n}\big[(1+\Delta)^{1+2n}(\varepsilon_e - \varepsilon_s)(n\varepsilon_f + (1+n)\varepsilon_s) + (\varepsilon_s - \varepsilon_f)(n\varepsilon_e + (1+n)\varepsilon_s)\big]}{n(1+n)(\varepsilon_f - \varepsilon_s)(\varepsilon_s - \varepsilon_e) + (1+\Delta)^{1+2n}(\varepsilon_e(1+n) + n\varepsilon_s)(n\varepsilon_f + (1+n)\varepsilon_s)} \tag{C.18c}$$

Further let us consider the rotated nanoparticle, which is far enough from the tip apex located at the point $\{0,0,h\}$. For this case $h \gg R + \Delta R$, and so only the first term ($n=1$) is significant in Eq.(C.18c).In this case the force components acquire a much simpler form

$$F_3 \approx \frac{-Q^* p_f}{2\pi\varepsilon_0 \varepsilon_e h^3} - \frac{Q^{*2} R^3}{2\pi\varepsilon_0 h^5} \frac{(1+\Delta)^3 \big[(1+\Delta)^3(\varepsilon_e - \varepsilon_s)(\varepsilon_f + 2\varepsilon_s) + (\varepsilon_s - \varepsilon_f)(\varepsilon_e + 2\varepsilon_s)\big]}{2(\varepsilon_f - \varepsilon_s)(\varepsilon_s - \varepsilon_e) + (1+\Delta)^3(2\varepsilon_e + \varepsilon_s)(\varepsilon_f + 2\varepsilon_s)}, \qquad F_{1,2} \approx \frac{Q^* p_{f1,2}}{4\pi\varepsilon_0 \varepsilon_e h^3} \approx 0. \tag{C.19}$$

By virtue of Newton's third law the same force, but with opposite sign, should act on the core-shell particle.

## APPENDIX D

The action of a Hadamard matrix $H = \frac{1}{\sqrt{2}}\begin{pmatrix} 1 & 1 \\ 1 & -1 \end{pmatrix}$ on a vector state $|A\rangle = (x, y)$ is the following: the matrix action rotates the vector by 45 degrees and provides a mirror reflection with respect to the y axis.

The action of a CNOT matrix $U_{CNOT} = \begin{pmatrix} 1 & 0 & 0 & 0 \\ 0 & 1 & 0 & 0 \\ 0 & 0 & 0 & 1 \\ 0 & 0 & 1 & 0 \end{pmatrix}$ leaves the upper block unchanged, and the bottom block becomes rotated by 90 degrees with a mirror reflection with respect to axis y. CNOT application on basis vectors leads to $|AA\rangle \rightarrow |AA\rangle$, $|AB\rangle \rightarrow |AB\rangle$, $|BA\rangle \rightarrow |AB\rangle$ and $|BB\rangle \rightarrow |BA\rangle$.